# New Horizons: Anticipated Scientific Investigations at the Pluto System

Leslie A. Young[1*], S. Alan Stern[1], Harold A. Weaver[2], Fran Bagenal[3], Richard P. Binzel[4], Bonnie Buratti[5], Andrew F. Cheng[2], Dale Cruikshank[6], G. Randall Gladstone[7], William M. Grundy[8], David P. Hinson[9], Mihaly Horanyi[3], Donald E. Jennings[10], Ivan R. Linscott[9], David J. McComas[7], William B. McKinnon[11], Ralph McNutt[2], Jeffery M. Moore[6], Scott Murchie[2], Carolyn C. Porco[12], Harold Reitsema[13], Dennis C. Reuter[10], John R. Spencer[1], David C. Slater[7], Darrell Strobel[14], Michael E. Summers[15], G. Leonard Tyler[9]

[1]*Southwest Research Institute, Boulder CO, U.S.A.*

[2]*Johns Hopkins University Applied Physics Lab., Laurel MD, U.S.A.*

[3]*University of Colorado, Boulder CO, U.S.A.*

[4]*Massachusetts Institute of Technology, Cambridge MA, U.S.A.*

[5]*Jet Propulsion Laboratory, Pasadena CA, U.S.A.*

[6]*NASA Ames Research Center, Moffett Field CA, U.S.A.*

[7]*Southwest Research Institute, San Antonio TX, U.S.A.*

[8]*Lowell Observatory, Flagstaff AZ, U.S.A.*

[9]*Stanford University, Stanford CA, U.S.A.*

[10]*NASA Goddard Space Flight Center, Greenbelt MD, U.S.A.*

[11]*Washington University, Saint Louis MO, U.S.A.*

[12]*Space Science Institute, Boulder CO, U.S.A.*

[13]*Ball Aerospace and Technologies Corporation, Boulder CO, U.S.A.*

[14]*Johns Hopkins University, Baltimore MD*

[15]*George Mason University, Fairfax, VA, U.S.A.*

(*Author for correspondence: email layoung@boulder.swri.edu)

# Abstract

The New Horizons spacecraft will achieve a wide range of measurement objectives at the Pluto system, including color and panchromatic maps, 1.25-2.50 micron spectral images for studying surface compositions, and measurements of Pluto's atmosphere (temperatures, composition, hazes, and the escape rate). Additional measurement objectives include topography, surface temperatures, and the solar wind interaction. The fulfillment of these measurement objectives will broaden our understanding of the Pluto system, such as the origin of the Pluto system, the processes operating on the surface, the volatile transport cycle, and the energetics and chemistry of the atmosphere. The mission, payload, and strawman observing sequences have been designed to acheive the NASA-specified measurement objectives and maximize the science return. The planned observations at the Pluto system will extend our knowledge of other objects formed by giant impact (such as the Earth-moon), other objects formed in the outer solar system (such as comets and other icy dwarf planets), other bodies with surfaces in vapor-pressure equilibrium (such as Triton and Mars), and other bodies with $N_2$:$CH_4$ atmospheres (such as Titan, Triton, and the early Earth).

# 1 Introduction

The New Horizons spacecraft was launched on 19 January 2006 on its 9.5-year journey to make the first reconnaissance of the Pluto system, including its satellites Charon, Nix, and Hydra, and a possible extended mission to one or more Kuiper Belt Objects (KBOs). While spacecraft have visited each of the terrestrial and jovian planets, this will be the first spacecraft to explore any of the ice dwarf worlds that dominate the third, outer portion of our solar system. By improving our knowledge of these distant bodies, we will extend our understanding of ice dwarfs and Kuiper Belt Objects (KBOs) and of the origin and evolution of our solar system.

New Horizons payload consists of Alice, an ultraviolet spectrometer; Ralph/MVIC, a visible panchromatic and color imager; Ralph/LEISA, an infrared imaging spectrometer; REX, a radio science experiment; LORRI, a high-resolution panchromatic imager; PEPSSI, an energetic particle detector, SWAP, a solar wind analyzer; and SDC, a student-built dust counter. Further detail is given in companion papers in this issue.

In this article, we describe the scientific investigations that the New Horizons science team (Table I) intends to perform at the Pluto system. In Section 2, we describe the science goals for which the New Horizons mission was designed. In Section 3, we describe the planned observations at Pluto and the Pluto system to fulfill these goals. In Section 4, we show the mapping of the observations to the specific mission goals, and in Section 5, we discuss some of the ways in which the planned observations will further planetary science in general.

TABLE I
New Horizons Science Team

| Name | Role |
|---|---|
| Alan Stern (SwRI) | New Horizons PI, Alice PI, Ralph PI |
| Hal Weaver (JHU/APL) | New Horizons PS, LORRI IS, GGI Team Member, ATM Team Member |
| Leslie Young (SwRI) | New Horizons Deputy PS, ATM Team Member |
| Fran Bagenal (CU) | P&P Lead, EPO lead |
| Richard Binzel (MIT) | GGI Team Member |
| Bonnie Buratti (JPL) | GGI Team Member |
| Andrew Cheng (JHU/APL) | LORRI PI, P&P Team Member |
| Dale Cruikshank (NASA/ARC) | COMP Team Member |
| Randall Gladstone (SwRI) | ATM Lead |
| William Grundy (Lowell) | COMP Lead |
| David Hinson (Stanford) | ATM Team Member |
| Mihaly Horanyi (CU) | SDC PI, P&P Team Member |
| Donald Jennings (NASA/GSFC) | Ralph/LEISA PI, COMP Team Member |
| Ivan Linscott (Stanford) | REX IS, ATM Team Member |
| Dave McComas (SwRI) | SWAP PI, P&P Team Member |
| William McKinnon (Wash. U) | GGI Team Member |
| Ralph McNutt (JHU/APL) | PEISSI PI, P&P Team Member, ATM Team Member |
| Jeffrey Moore (NASA/ARC) | GGI Lead |
| Scott Murchie (JHU/APL) | GGI Team Member |
| Carolyn Porco (SSI) | GGI Team Member |
| Harold Reitsema (Ball) | GGI Team Member |
| Dennis Reuter (NASA/GSFC) | Ralph IS, also GGI, COMP Team Member |
| David Slater (SwRI) | Alice IS, ATM Team Member |
| John Spencer (SwRI) | GGI Deputy, also COMP Team Member |
| Darrell Strobel (JHU) | ATM Team Member |
| Michael Summers (GMU) | ATM Team Member |
| Leonard Tyler (Stanford) | REX PI, ATM Team Member |

PI = Principal Investigator. PS = Project Scientist. IS = Instrument Scientist. ATM = Atmospheres Theme Team. GGI = Geology, Geophysics and Imaging Theme Team. COMP = Composition Theme Team. P&P = Particles and Plasmas Theme Team.

# 2. Mission Science Goals

The Pluto system (Table II) consists of Pluto, discovered in 1930, its large satellite Charon, discovered in 1978, and two smaller moons Nix and Hydra, discovered in 2005. Pluto's heliocentric orbit has an eccentricity of 0.25, with a heliocentric distance that ranges from 29.7 AU at perihelion (1989 Sep 5) to 49.3 AU (2114 Feb 19), leading to a factor of 2.8 in received insolation over its orbit. Pluto has a high obliquity of 119 deg (Tholen and Buie 1997): the sub-solar latitude ranges from 0 deg at equinox (1987 Dec 11) to 57.5 deg at the solstices (-57.5 deg on 2029 Jul 10). Pluto will be 32.9 AU from the Sun with a sub-solar latitude of -49.5 degrees at the time of the New Horizons closest approach. (Here and in the remainder of the paper, as is convention for the New Horizons mission, we use the IAU definition of Pluto's North pole, which is equivalent to Pluto's south rotational pole.)

Pluto and Charon form a double planet, with the center of mass of the system exterior to Pluto. Despite their proximity, they are vastly different worlds (Stern 1992). Pluto is red in color, with a surface covered with volatiles ($N_2$, $CH_4$, CO) that support a seasonally variable atmosphere. The sublimation/condensation cycle, plus photochemical processing, leads to an overall high albedo, and large albedo variations. Charon, in contrast, is neutral in color, and has a water-dominated surface with little spatial variation; no Charonian atmosphere has been detected. Little is currently known about the physical properties of Nix and Hydra beyond their orbits, magnitudes, and gross colors (Buie et al. 2006, Stern et al. 2006), with sizes and masses that have been inferred from likely ranges of albedos and densities.

TABLE II
Inventory of the Pluto system

| Object | Semi-major axis, *a* (km)[a] | Absolute magnitude, *H*[c] | Radius (km) | Mass ($10^{22}$ kg) | Geometric Albedo (V) | Surface Constituents | Atmosphere |
|---|---|---|---|---|---|---|---|
| Pluto | 2042±86[b] | -0.5368 ± 0.0022[d] | 1140-1190[e] | 1.305±0.006[d] | 0.52[h] | $N_2$:$CH_4$, $CH_4$, CO, possible tholins[i] | $N_2$, $CH_4$, CO[k] |
| Charon | 17529±86[b] | 1.3488 ± 0.0868[d] | 606.0 ± 1.5[f] | 0.152±0.006[d] | 0.35[h] | $H_2O$, possible $NH_3$, $NH_3$• 2 $H_2O$[j] | None detected[f] |
| Nix (2005/P2) | 48675 ± 121[b] | 8.636 ± 0.095[b] | 50-130[g] | <0.0004[g] | 0.04-0.35[g] | Unknown | Unlikely |
| Hydra (2005/P1) | 64780 ± 88[b] | 8.483 ± 0.089[b] | 60-165[g] | <0.0004[g] | 0.04-0.35[g] | Unknown | Unlikely |

[a] semi-major axis of orbit, relative to system barycenter. Charon semimajor axis (relative to Pluto) is 19571.4 ± 4.0 (Buie et al. 2006). [b]Buie et al. 2006. [c]V magnitude at geocentric distance = 1 AU, heliocentric distance = 1 AU, phase angle = 0 degrees. [d]Average over a rotation period from observations in 1992 and 1993 (Buie et al. 1997a). [e]Tholen and Buie 1997. [f]Young et al. 2005, Gulbis et al. 2006, Sicardy et al. 2006. [g]Weaver et al. 2006. Albedo ranges and densities are assumed. [h]Based on rotationally averaged magnitudes. Radius of 1180 km used for Pluto albedo. [i]Owen et al. 1993, Olkin et al. 2006b. [j]Brown & Calvin 2000, Brown 2002. [k]Owen et al. 1993, Young et al 1997, Young et al. 2001b.

Based on the remarkable growth in our knowledge of Pluto and Charon in the 1980's and early 1990's, and guided by the Voyager encounter of Triton, in 1992 NASA's Outer Planet Science Working Groups (OPSWG, S. A. Stern, chair) described three high priority, nine second priority, and four third priority science objectives for Pluto system flyby reconnaissance. These objectives were adopted, essentially verbatim, in the October 1995 Pluto-Kuiper Express Science Definition Team Report (J. I. Lunine, chair), the September 1999 Announcement of Opportunity for the Pluto-Kuiper Express (PKE-MPD AO: 99-OSS-04), and the January 2001 Announcement of Opportunity for the Pluto-Kuiper Belt Mission (PKB-AO: AO:01-OSS-01) (Table III). More detail on the history of the mission is given by Stern, this issue.

Important new results for Pluto and Charon since the Mission Science Goals were first proposed include the discovery of the moons Nix and Hydra (Weaver et al. 2006), the measurement of Pluto's post-perihelion atmospheric expansion (Elliot et al. 2003, Sicardy et al. 2003), new limits on Charon's atmosphere and measurement of its radius (Gulbis et al. 2006, Person et al. 2006), rotationally or spatially resolved measurements of Pluto's temperature (Lellouch et al 2000), surface composition (Grundy and Buie 2001), and color (Young et al. 2001a). These have upheld, rather than challenged, the importance of the science goals to which the New Horizons mission was designed.

TABLE III
New Horizons Pluto System Science Objectives
as stated in PKB-AO: AO:01-OSS-01

| Group 1 (required) | |
|---|---|
| Goal No. | Goal |
| **1.1** | **Characterize the global geology and morphology of Pluto and Charon** |
| 1.1a | *Hemispheric panchromatic maps:* Obtain panchromatic viewable disk coverage of both Pluto and Charon at a resolution of 1 kilometer per line pair (1 km/lp), or equivalent. Viewable disk means the entire lit and visible surface of the target body viewed from the spacecraft at a single point in time during the approach to the target. The 1 km/lp objective applies to the subspacecraft point; it is understood that a combination of image projection effects and spacecraft data storage limitations may degrade resolution away from the subspacecraft point. |
| 1.1b | *Color maps:* Obtain viewable disk coverage of both Pluto and Charon in 2 to 5 color bands at a resolution of 3-10 km/lp (or equivalent). The resolution objective applies to the subspacecraft point; it is understood that a combination of image projection effects and spacecraft data storage limitations may degrade resolution away from the subspacecraft point. |
| 1.1c | *Phase angle coverage:* Obtain sufficient imaging at moderate and high phase angles to specify the phase integrals of Pluto and Charon. |
| 1.1d | *Image dynamic range and signal-to-noise ratio (S/N):* For all imaging, provide sufficient dynamic range to cover brightness contrasts of up to 30 (i.e., normal albedo between 0.03 and 1) with an average S/N goal of about 100, but somewhat lower S/N in the darkest regions. |
| **1.2** | **Map surface composition of Pluto and Charon** |
| 1.2a | *Hemispheric infrared spectroscopic maps:* Obtain infrared spectroscopic maps of one hemisphere of both Pluto and Charon with approximately 10 km/pixel resolution at disk center with the ability to detect a <0.02 change in albedo everywhere in the spectrum. |
| 1.2b | *Goal for compositional determination:* Determine the spatial distribution of frozen $N_2$ and secondary constituents such as CO, $CH_4$. Determine quantitatively the presence of such additional major exposed volatiles, hydrocarbons, and minerals (or rocks) as may exist, all at spatial resolution of 5-10 km/pixel or equivalent. |
| 1.2c | *Spectral coverage and resolution:* For each spatial resolution element, obtain a spectral resolution ($\lambda/\Delta\lambda$) of at least 250 over all or part of the 1 - 5 micron region (or beyond, if relevant). |
| **1.3** | **Characterize the neutral atmosphere of Pluto and its escape rate.** |
| 1.3a | *Composition*: Determine the mole fractions of $N_2$, CO, $CH_4$ and Ar in Pluto's atmosphere to at least the 1% level of the total mixing ratio. |
| 1.3b | *Thermospheric thermal structure:* Measure T and dT/dz at 100 km vertical resolution to 10% accuracy at gas densities of $10^9$ cm$^{-3}$ and higher. |
| 1.3c | *Aerosols*: Characterize the optical depth and distribution of near-surface haze layers over Pluto's limb at a vertical resolution of 5 km or better. |
| 1.3d | *Lower atmospheric thermal structure:* Measure temperature and pressure at the base of the atmosphere to accuracies of ± 1 K and 0.1 microbar. |
| 1.3e | *Evolution*: Determine the escape rate. |

TABLE III (continued)
New Horizons Pluto System Science Objectives
as stated in PKB-AO: AO:01-OSS-01

| Group 2 (strongly desired) | |
|---|---|
| Goal No. | Goal |
| 2.1 | Characterize the time variability of Pluto's surface and atmosphere |
| 2.2 | Image Pluto and Charon in stereo |
| 2.3 | Map the terminators of Pluto and Charon with high resolution |
| 2.4 | Map the surface composition of selected areas of Pluto and Charon with high resolution |
| 2.5 | Characterize Pluto's ionosphere and solar wind interaction |
| 2.6 | Search for neutral species including H, $H_2$, HCN, and $C_xH_y$, and other hydrocarbons and nitriles in Pluto's upper atmosphere, and obtain isotopic discrimination where possible |
| 2.7 | Search for an atmosphere around Charon |
| 2.8 | Determine bolometric Bond albedos for Pluto and Charon |
| 2.9 | Map the surface temperatures of Pluto and Charon. |
| Group 3 (desired) | |
| Goal No. | Goal |
| 3.1 | Characterize the energetic particle environment of Pluto and Charon |
| 3.2 | Refine bulk parameters (radii, masses, densities) and orbits of Pluto and Charon |
| 3.3 | Search for magnetic fields of Pluto and Charon (indirectly addressed by New Horizons) |
| 3.4 | Search for additional satellites and rings. |

## 2.1 Geology and Morphology Goals (Goals 1.1, 2.1, 2.2, 2.3, 2.8)

From rotational photometric and compositional lightcurves, mutual events (Pluto and Charon transiting and eclipsing each other), and the first post-repair HST images, it is clear that Pluto had extreme albedo variations, suggesting a complex geological and geomorphological surface (Buie et al, 1997b, Stern et al. 1997, Young et al. 1999, 2001a; see Fig. 1). Panchromatic images of Pluto, Charon, Nix, and Hydra will provide a spatial context for the interpretation of New Horizons and observatory-based (ground-based or near-Earth space-based) data. Broad-band colors can be used with albedo to differentiate the surfaces into distinct color/albedo units, presumably related to geological units, as was done for Triton (Brown et al. 1995). On Pluto, global albedo maps are also critical for understanding the seasonal volatile transport cycle, since the temperatures of both volatiles and substrates depend on their bond albedos. While the requirements (Goals 1.1a and 1.1b) are to image one hemisphere of Pluto and Charon, the New Horizons mission will map as much of the surfaces of Pluto, Charon, Nix and Hydra as possible. Because the frosts on Pluto's surface are expected to migrate, New Horizons will also image Pluto over time (Goal 2.1).

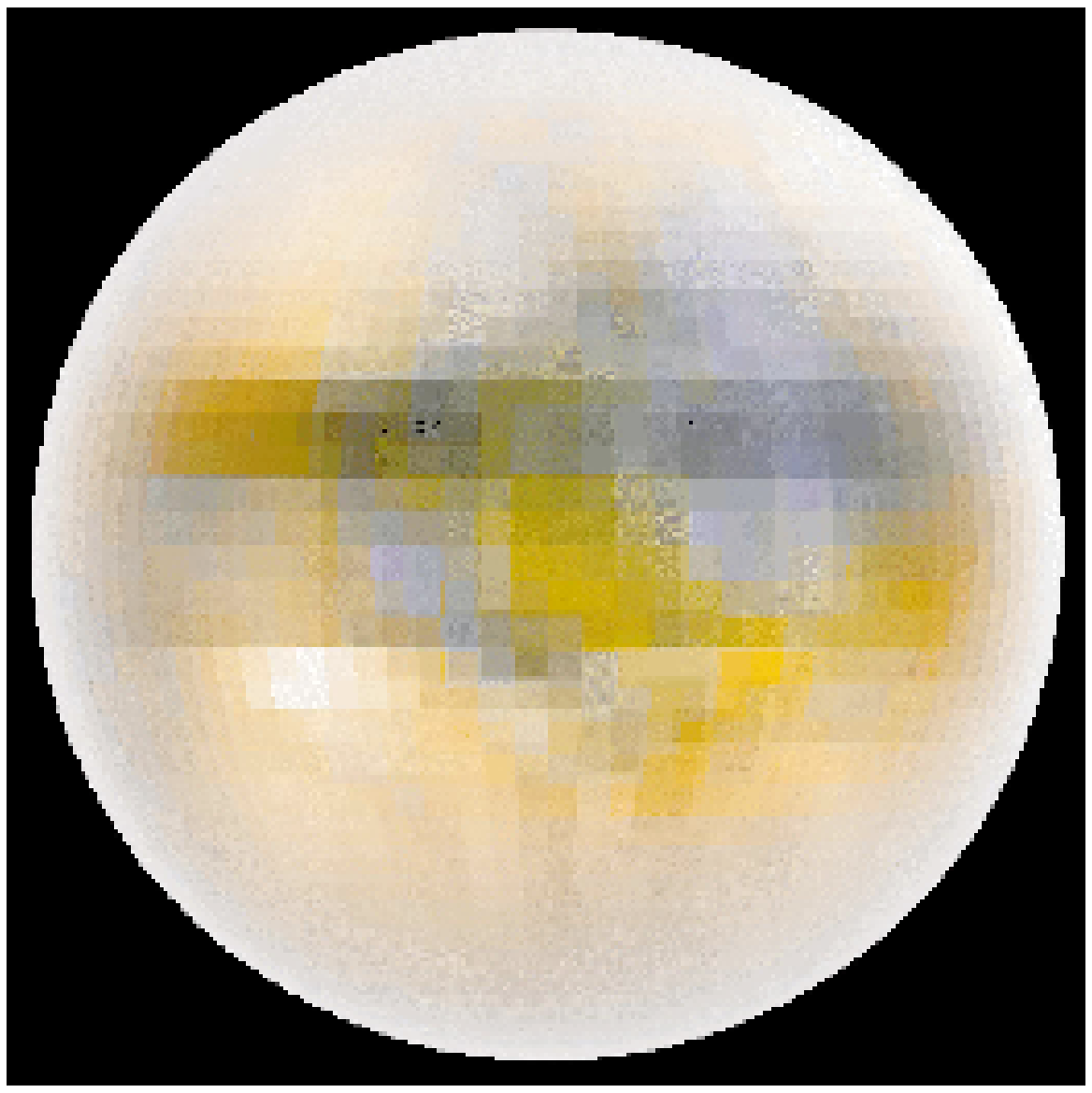

Fig 1. False-color B-V image of Pluto (Young et al. 2001a) rotated so that the IAU North pole (rotational South pole, and current winter pole) is up.

Both hemispheric maps (Goal 1.1) and maps at higher resolution (Goal 2.3) will elucidate the underlying geological and geomorphological processes. For example, visible imaging will detect craters, tectonic features, and changes in reflectivity indicative of various compositions, origins, and evolution, while observations of the same areas at different viewing geometries can yield bond albedo (Goal 2.8, related to the phase integrals, Goal 1.1.c), and grain size. Height and terrain information are also important, and can be derived from stereo imaging (Goal 2.2) and from shadows and shading (especially effective at high phase angle, Goal 2.3, and combined with a photometric phase function, Goal 1.1c).

New Horizons flies two cameras for panchromatic imaging (LORRI and Ralph/MVIC), one of which (Ralph/MVIC) also has three broad color filters and a filter designed to detect $CH_4$ ice.

## 2.2 Surface Composition Goals (Goals 1.2, 2.1, 2.4, 2.9)

Infrared spectra of surfaces in the outer solar system (Goal 1.2c) are rich in spectral features. These spectra are particularly diagnostic shortward of 2.5 microns, because sunlight (and consequently signal-to-noise) drops dramatically at longer wavelengths. Spectroscopy of Pluto's surface shows spectral features of frozen $N_2$ (2.15 micron), $CH_4$ (weak and strong bands that dominate Pluto's spectrum) and CO (1.58 and 2.35 micron) (Owen et al. 1993, Douté et al. 1999, Cruikshank et al. 1997; see Fig 2a). Charon's infrared spectrum shows $H_2O$ ice (including the 1.65 micron feature indicative of crystalline $H_2O$ ice) and probably $NH_3$ or its hydrate $NH_3$•2 $H_2O$ near 2.2 micron (Brown and Calvin 2000, Brown 2002), although other species, such as aluminum-bearing phyllosilicates, also have absorption bands at 2.215 micron (Buie and Grundy 2000).

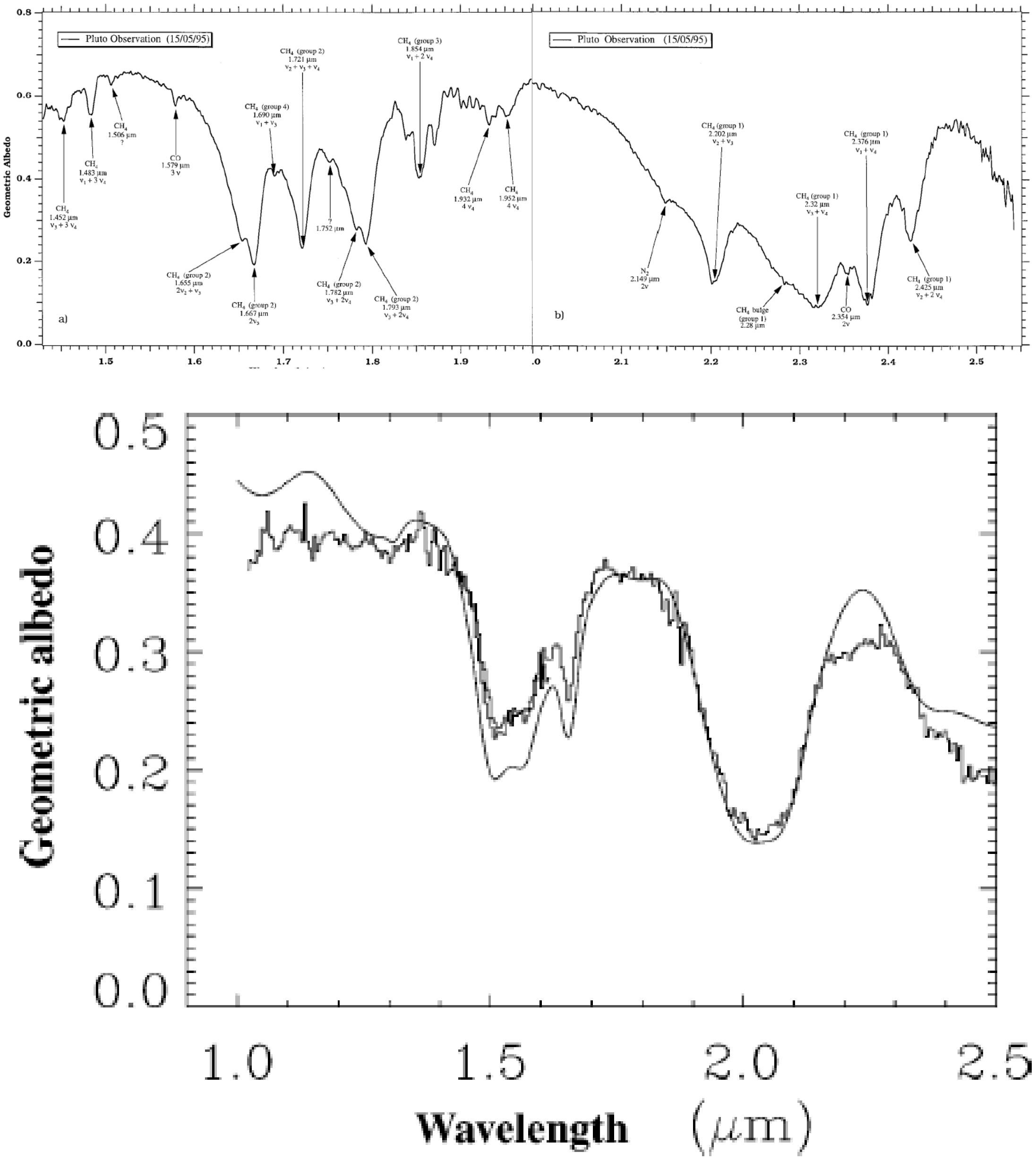

Fig. 2. Top: Pluto's near-infrared spectrum (Douté et al. 1999), showing absorption bands due to $N_2$, $CH_4$, and CO ices. Bottom: Charon's near-infrared spectrum (Brown 2002), with the best-fit water ice model as a smooth line, showing crystalline $H_2O$ ice and an additional absorber at 2.2 micron (most likely $NH_3$ or its hydrate).

$N_2$, $CH_4$, and CO are all volatile at Pluto's surface temperature, and therefore support Pluto's atmosphere and participate in Pluto's seasonal cycle (Spencer et al. 1997). These ices are not uniformly distributed over Pluto's surface (Grundy and Buie 2001). An understanding of the spatial distribution of frozen $N_2$, $CH_4$, and CO (Goal 1.2b) is vital for understanding the surface energy balance, since energy is transported across Pluto's surface through latent heat of sublimation. Spatially resolved spectra will also allow a more sensitive search for additional surface species (Goal 1.2b). On Pluto, for example, exposed $H_2O$ ice can be more easily identified without the interference of $CH_4$ bands (Grundy and Buie 2002). On Charon, $NH_3$ may be confirmed in spatially resolved spectra by the detection of the 1.99 micron band of $NH_3$, which may be suppressed by the presence of the broad $H_2O$ band at 2.0 micron in disk-averaged spectra (Buie and Grundy 2000).

With sufficient signal-to-noise ratio and spectral resolution (Goals 1.2a and 1.2.c), infrared (IR) spectra can be used for much more than the mere determination of the presence of surface species. The IR spectra of $N_2$, $CH_4$, and $H_2O$ are all temperature dependent (Goal 2.9; Tryka et al. 1995, Grundy et al. 1993, 1999, 2002), especially when the simultaneous observation of both strong and weak bands of $CH_4$ can break the grainsize/temperature ambiguity. The bands of frozen $CH_4$ shift according to whether or not the $CH_4$ ice is pure or in a solid solution (i.e., whether a $CH_4$ molecule neighbors other $CH_4$ molecules or different host, or matrix, molecules).

Observations over a long timebase may detect the migration of volatiles (Goal 2.1). High spatial resolution composition maps (Goal 2.9) allow the connection between geologic processes and composition, illuminating questions such as: what are the chromophores causing Pluto's red color? If New Horizons finds sublimations scarps, will they be associated with areas of volatiles? Do craters reveal a change in composition with depth?

For composition mapping, New Horizons flies a 1.25-2.5 micron spectral imager (Ralph/LEISA), augmented by mapping of $CH_4$ through its visible-wavelength 0.89 micron band (Ralph/MVIC).

## 2.3 Atmospheric and Particles/Plasma Goals (Goals 1.3, 2.1, 2.5, 2.6, 2.7, 3.1)

Pluto's atmosphere is believed to be in vapor pressure with its surface ices. Of the three volatiles detected on Pluto ($N_2$, CO, and $CH_4$), $N_2$ has the highest vapor pressure, and so dominates atmosphere, with $CH_4$ and CO as trace components (Owen et al. 1993). Although the total column abundance of $CH_4$ has been measured (Young et al. 1997), its vertical profile and mixing ratio are unknown. Only upper limits on gaseous CO exist (Young et al. 2001b; Bockelée-Morvan et al. 2001). Models of Pluto's atmosphere (Yelle and Elliot 1997, Summers and Strobel 1997, Krasnopolsky and Cruikshank; see Fig 3) depend critically on the bulk composition (Goal 1.3a) for energetics and thermal structure (Goal 1.3b), chemistry (Goal 2.6), condensation products (Goal 1.3c), escape rates (Goal 1.3e), and the formation and composition of an ionosphere (Goal 2.5).

Pluto has an atmosphere that is thick enough to globally transport volatiles (e.g., Yelle and Elliot 1997). In theory, the surface pressure is a simple function of $N_2$ ice temperature. In practice, even if surface temperatures were measured with perfect accuracy, much is uncertain about the relation between the $N_2$ surface pressure (Goal 1.3d) and the temperature of N2:CH4 ice (Trafton et al. 1998), since the less-volatile $CH_4$ may form a barrier to $N_2$ sublimation, or a near-surface turbulent layer may affect the vertical transport of energy, $CH_4$, and CO, affecting the atmospheric composition (Goal 1.3a) and near-surface temperature structure (Goal 1.3d).

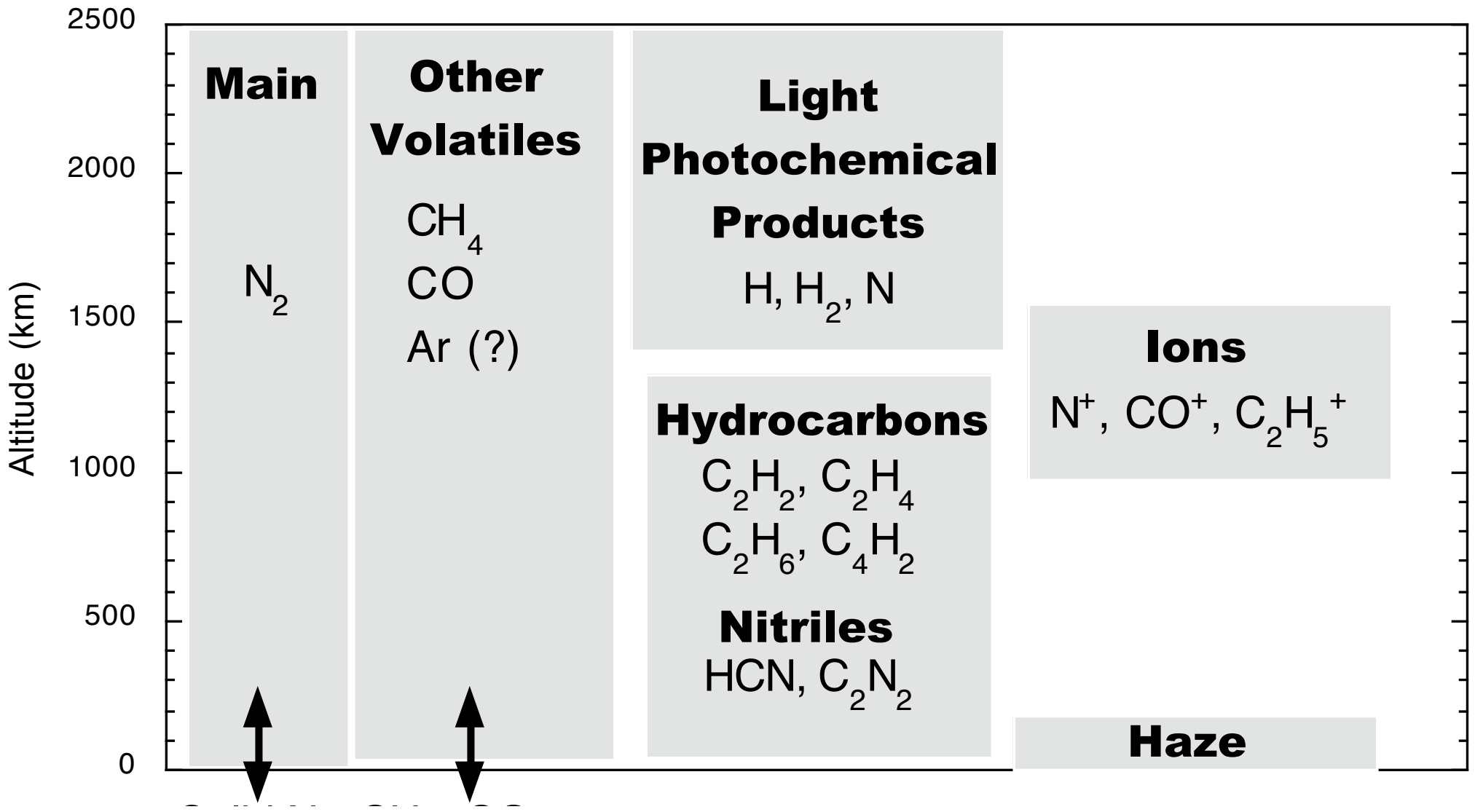

*Figure 3.* Summary of atmospheric structure and composition, based on the models of Summers et al. 1997.

Pluto's atmosphere is likely to be in hydrodynamic escape (Goal 1.3e) with escape rates of a few $10^{10}$ cm$^{-2}$ s$^{-1}$ (Trafton et al. 1997, Krasnopolsky 1999), equivalent to several km of frost over the age of the solar system. Light photolysis products (H, $H_2$, N) escape, and heavier products may escape at higher $N_2$ flux rates (Trafton et al. 1997). Thus, Pluto's escaping atmosphere should leave its mark on Pluto's surface in the sublimation record, and possibly the amount of photochemical products that have rained onto the surface. Pluto's atmosphere is expected to extend past the orbit of Charon, although a secondary atmosphere at Charon is not expected (Goal 2.7). Pluto's ionosphere and extended atmosphere should interact with the solar wind (Goal 2.5), with high fluxes leading to a large region (tens of Pluto radii) of comet-like interaction, and low fluxes leading to a smaller region with Venus-like interaction (Bagenal et al. 1997). Neutrals escaping from Pluto's atmosphere are ionized by solar EUV photons or by charge transfer, leading to an energetic particle environment of pickup ions (Goal 3.1).

New Horizons addresses the atmospheric goals primarily through UV occultations and airglow observations (Alice) and uplink radio occultations (REX). The haze observations are made by imaging (Ralph/MVIC, LORRI) and UV occultations, with SWAP and PEPSSI addressing the solar wind, energetic particle, and atmospheric escape rate goals.

## 2.4 Interiors, environment, origin and evolution (Goals 3.2, 3.3, 3.4)

Clues to the formation and evolution of Pluto, Charon, Nix, and Hydra (presumably by giant impact, Stern et al. 1997, Canup 2005) lie in their orbits (Dobrovolskis et al. 1997; Malhotra and Williams 1997; Goal 3.2) and bulk properties (Tholen and Buie 1997; Goal 3.2). Further information can be found from the discovery of (or limits on) additional rings or satellites (Goal 3.4) interior to the Hill sphere (6.0 x $10^6$ km; Steffl et al. 2007). The heliocentric orbit of Pluto points to its formation in the outer solar nebula and its subsequent resonance capture by Neptune. The obliquity of each body in the system, the orbits of each body about the system barycenter, the state of tidal lock, and the interior structure of each body give additional constraints on the formation of the system. The interior structure gives external expression in the masses, radii, and shapes (and, possibly, the magnetic fields, Goal 3.3) of each body.

New Horizons will measure the orbits and bulk properties by a combination of imaging (LORRI, Ralph/MVIC) and the radio system (REX). If Pluto has a substantial magnetic field, this may be detectable by its effect on the solar wind (SWAP). Finally, SDC will measure the dust environment of Pluto.

### 2.5 Kuiper belt observations

The Kuiper belt has emerged in recent years as a fundamental architectural element of the outer solar system. In addition to containing a rich reservoir of nearly pristine relics from the era of planet formation, the Kuiper belt provides context for understanding the origin and evolutionary environment of the Pluto-Charon system (e.g., Stern et al. 1997). The Kuiper Belt and scattered disk are currently considered the source region of the short-period Jupiter family comets (Farinella et al. 2000; Weissman & Levison 1997) and provide a link between our solar system and the debris disks seen around main sequence stars such as ß Pic. A goal of New Horizons is to investigate one or more Kuiper belt object (KBO) in an extended mission, fulfilling as many of the Group 1, 2 and 3 goals as possible.

## 3. Planned Observations of the Pluto System

### 3.1 Instrument Suite

The New Horizons payload (Table IV) addresses all of the Group 1, 2, and 3 goals listed in Table III (except the Group 3 goal of a search for magnetic fields, which is only addressed indirectly). The payload as a whole is described in Weaver et al. (this issue), with the individual instruments described in Cheng et al., Horanyi et al., McNutt et al., McComas et al., Reuter et al., Stern et al., and Tyler et al., this issue.

TABLE IV
New Horizons Payload

| Instrument | Description | Comments |
|---|---|---|
| Ralph/MVIC (Multi-spectral Visible Imaging Camera) | Visible panchromatic and color imager | Panchromatic and color imaging |
| Ralph/LEISA (Linear Etalon Imaging Spectral Array) | IR imaging spectrometer | Surface composition maps |
| Alice (not an acronym) | UV spectrometer | Upper atmospheric temperatures, composition and escape rate |
| REX (Radio EXperiment) | Radio science experiment (uplink and radiometry) | Lower atmospheric temperature, pressure, and density; radiometry |
| LORRI (LOng Range Reconnaissance Imager) | High-resolution panchromatic imager | Panchromatic imaging |
| SWAP (Solar Wind Around Pluto) | Solar wind analyzer | Solar wind interaction |
| PEPSSI (Pluto Energetic Particle Spectrometer Science Investigation) | Energetic particle detector | Energetic particle environment |
| SDC (Student Dust Counter)[a] | Dust counter | Interplanetary dust environment |

[a] Renamed VBSDC (Venetia Burney Student Dust Counter) after launch.

### *3.1.1 Ralph: MVIC and LEISA*

Ralph and its two focal planes, MVIC and LEISA, are summarized in Weaver et al. and described in detail in Reuter et al., this issue. MVIC's prime purpose is to obtain the Group 1 panchromatic and color maps. It also supports the Group 1 goal of characterizing of haze properties, the Group 1 composition maps, and the Group 2 and 3 goals of time variability, stereo imaging, terminator imaging, bond albedos, bulk parameters, and satellite/ring search. Both MVIC and LORRI take panchromatic images, with complementary observing techniques. Since MVIC operates its time-delay integration (TDI) panchromatic array in a scanning mode, it can obtain the Group 1 0.5 km/pixel hemispheric map of Pluto at a distance of 25,000 km in a single scan, observing the whole of Pluto's visible hemisphere in roughly 60 seconds without seams, and a change in scale of only ~3%. Furthermore, MVIC was specifically designed to take observations of Pluto's atmosphere after closest approach, to search for hazes in forward scattering.

MVIC three broad-band colors, blue (400-550 nm), red (540-700 nm), and near-IR (780-975 nm) were chosen to give information on color slopes, as well as atmospheric properties. The narrow band $CH_4$ filter (860-910 nm) allows mapping of $CH_4$ abundance through the well known 0.89 micron $CH_4$ feature seen on Pluto's surface, the strongest methane feature at CCD wavelengths. The NIR filter acts as the continuum comparison for this methane mapping. The 700-780 nm gap between the red and NIR includes another $CH_4$ band at 740 nm; combining information from panchromatic, blue, red, and NIR filters can give some information about band depth in this "virtual" filter.

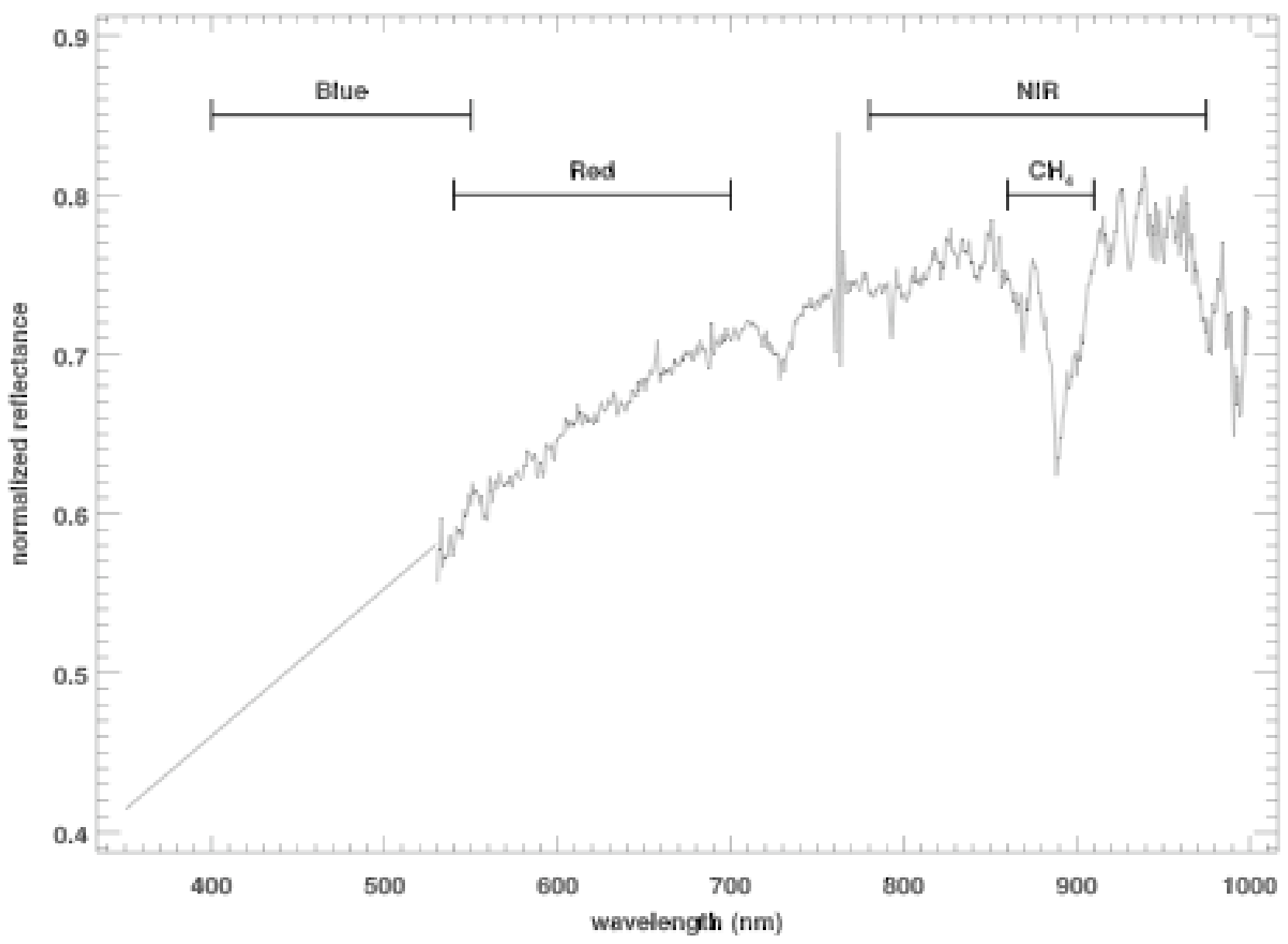

*Figure 4*. Spectrum of Pluto. (Grundy 1995, Buie et al. 1997a), and the MVIC color filters for comparison.

LEISA's prime purpose is to obtain the Group 1 surface composition maps, and to support the Group 2 and 3 goals of time variability, high-resolution surface composition, and surface temperatures. LEISA's full range of 1.25-2.5 micron with a spectral resolution (λ/Δλ) of roughly 250 was chosen to include absorption features of species previously detected on Pluto ($N_2$, $CH_4$, and CO) and Charon ($H_2O$, $NH_3$) as well as other species relevant to the outer solar system (Fig. 5). Because of the importance of the $N_2$ feature and 2.15 microns, as well as the $\nu_2 + \nu_3$ $CH_4$ band at 2.20 microns (diagnostic of pure vs. diluted $CH_4$ abundances; Quirico & Schmitt 1997), LEISA also measures the region covering 2.10-2.25 micron with a spectral resolution of roughly 560. LEISA operates by scanning its 256 x 256 pixel array across a target to build up a complete spectral image (Reuter et al., this issue).

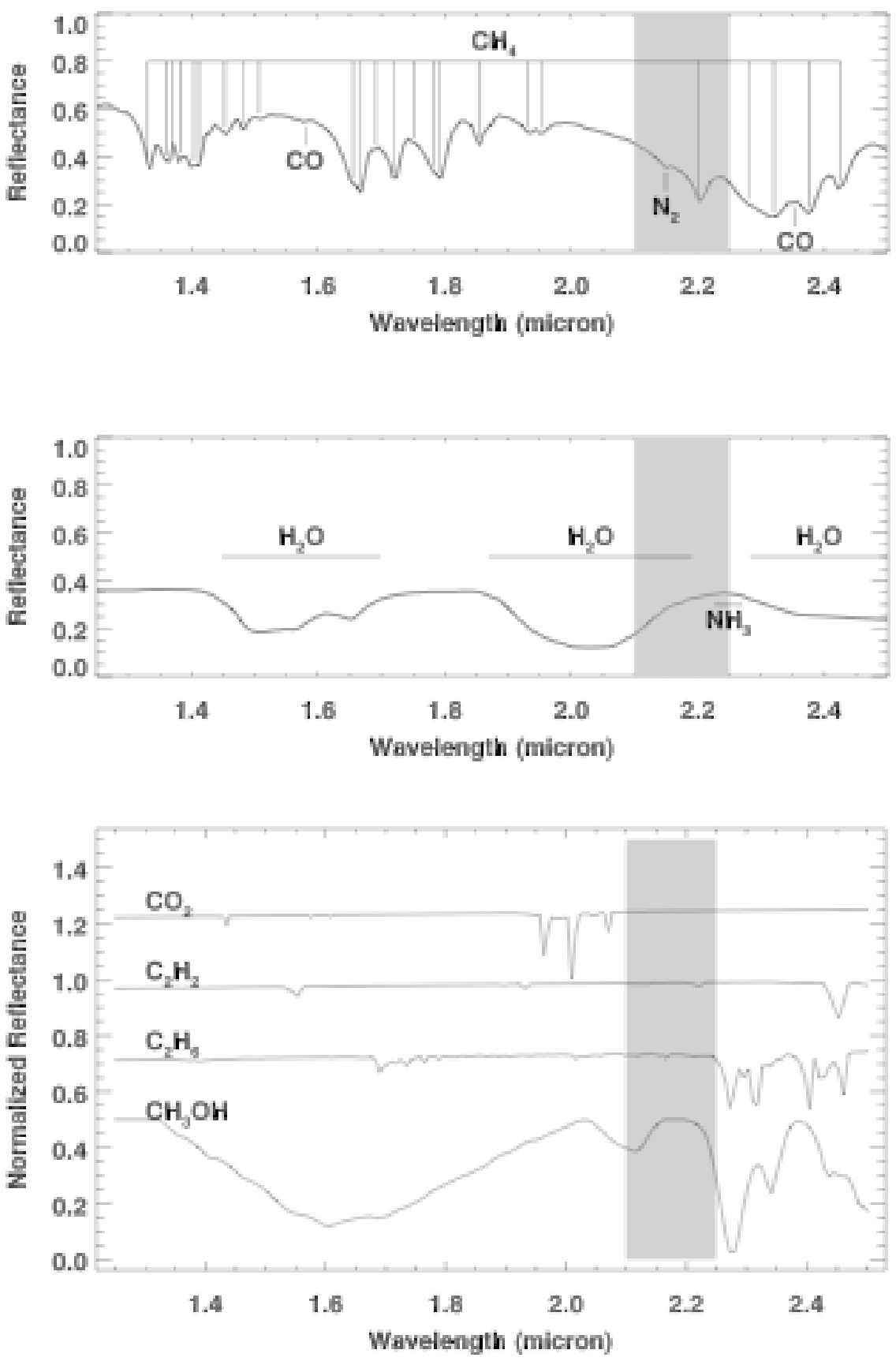

*Figure 5.* Spectrum of Pluto (top), Charon (middle), and other significant solids (bottom) over the LEISA spectral range, at a spectral resolution of 250 over 1.25-2.50 micron, and a resolution of 560 over 2.10-2.25 micron (indicated in gray).

### *3.1.2 Alice*

Alice is summarized in Weaver et al. and described in detail in Stern et al., this issue. Alice's prime purpose is to obtain the Group 1 atmospheric goals of composition, thermospheric temperature structure, aerosols, and evolution (escape rate). It also supports the Group 2 and 3 goals of time variability, minor atmospheric species, and Charon atmosphere search. Alice observes occultations of stars or the sun in a time-tagged mode with a spectral resolution of 3.5 Å. Longer integrations, for airglow or surface studies, are made in histogram mode. These can either be observed through Alice's 2 x 2 deg "box" portion of the slit or through the 0.1 x 2 deg "slot" portion, at resolutions of 3-4.5 Å for point sources or 9.0 Å for objects that fill the 0.1 degree x 4 degree "slot." The spatial resolution is 0.27 deg pixel$^{-1}$ along the slit. Alice's 465-1881 Å bandpass includes the electronic cutoffs of $N_2$, Ar, $CH_4$, and other hydrocarbons and nitriles that may be detectable by solar and stellar occultations (Fig. 6), and emission lines of CO, atomic H, Ar, and Ne that may be detectable as airglow (Fig 7). Because $N_2$ dominates Pluto's atmosphere, the thermal profile can be derived from the $N_2$ density profile. The temperatures at high altitudes serve as lower boundary conditions for atmospheric escape; Alice can also map the cloud of escaping H by Ly-$\alpha$ against the interstellar background.

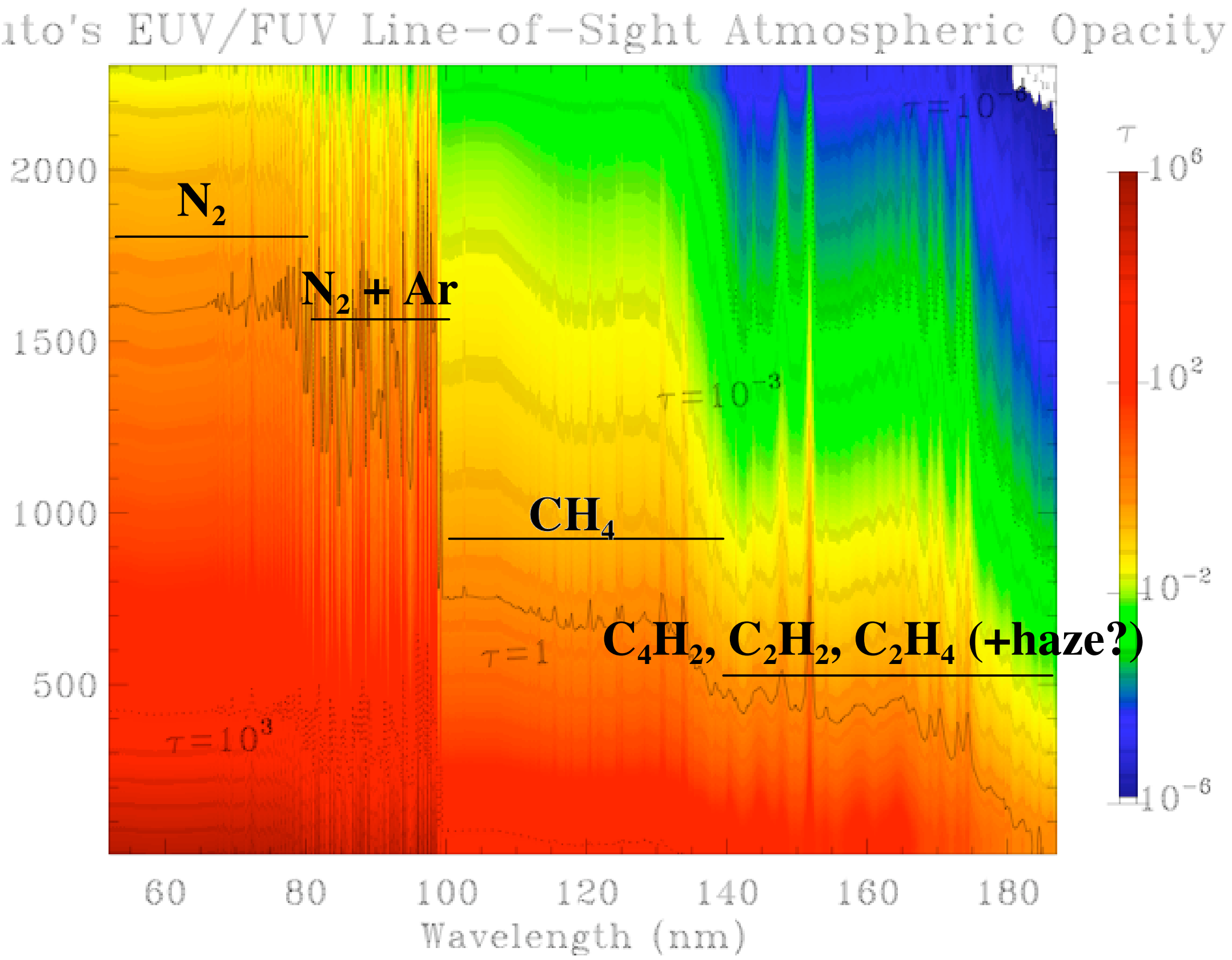

*Figure 6.* Simulated line-of-sight optical depth ($\tau$) for the M2 model atmosphere of Krasnopolsky and Cruikshank (1999), with an added 5% Ar.

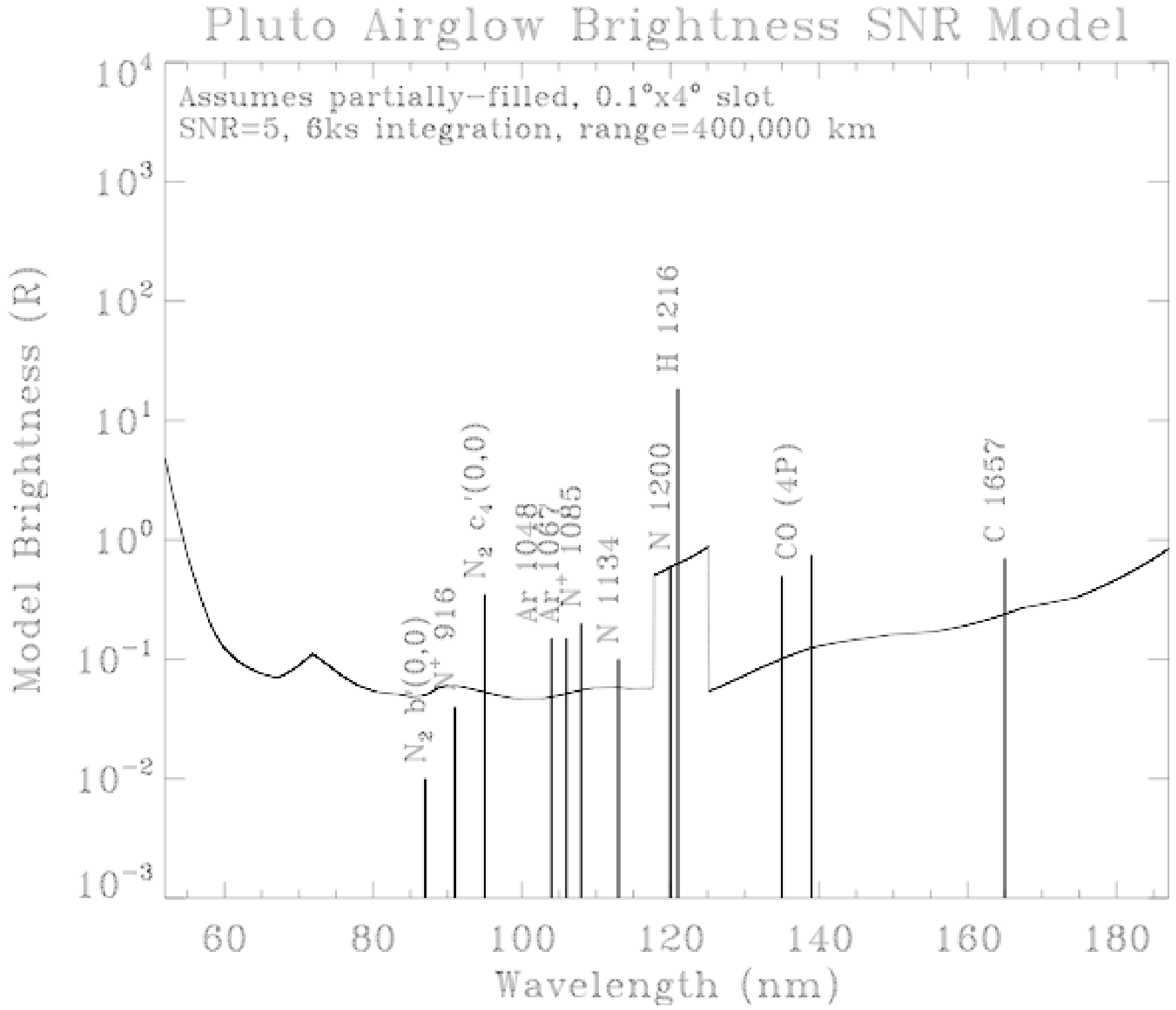

*Figure 7.* Some of the airglow lines in the Alice bandpass.

### *3.1.3 REX*

REX is summarized in Weaver et al. and described in detail in Tyler et al., this issue. REX's prime purpose is to obtain the Group 1 atmospheric goals of measuring the lower atmospheric pressure and temperature; during Earth occultations, REX will detect the delay of radio signals uplinked from multiple DSN antennas. REX will also address Group 2 and Group 3 objectives, including probing Pluto's electron density within its ionosphere, searching for Charon's atmosphere, refining bulk parameters like mass and radius, and measuring surface emission at 4.2 cm.

### *3.1.4 LORRI*

LORRI is summarized in Weaver et al. and described in detail in Cheng et al., this issue. LORRI augments most of the MVIC panchromatic imaging goals, including the Group 1 panchromatic mapping goals. Compared to MVIC, LORRI has a larger aperture, a square field of view (1024x1024 vs. MVIC's 5000 x 32 framing array), and finer angular resolution (5 microrad pixel$^{-1}$ vs. MVIC's 20 microrad pixel$^{-1}$). This makes LORRI especially valuable for long time-base observations of the Pluto system, for observing the "far side" hemisphere of Pluto during the last 6.4 days (one Pluto rotation) before closest approach, making observations where geometric fidelity is critical (which favors LORRI's stare mode over MVIC's scanning operation), and taking the highest resolution images near closest approach.

### 3.1.5 PEPSSI

PEPSSI is summarized in Weaver et al. and described in detail in McNutt et al., this issue. PEPSSI's prime purpose is the Group 3 goal of characterizing the energetic particle environment. In particular, PEPSSI's primary objective is to determine the mass, energy spectra, and directional distributions of these energetic particles, with a resolution able to discriminate between the various types of species expected at Pluto. The bulk of the energetic particles are expected to be pickup ions, or particles from Pluto's tenuous cloud of escaping neutrals, some billion of km in extent, that are ionized by solar EUV photons or charge transfer collisions with solar wind protons (Bagenal et al. 1997). These new ions then feel and react to the solar wind, following roughly cycloidal trajectories. In Pluto's frame of reference, these ions reach a maximum speed of twice the solar wind velocity, so that ions have a maximum energy of $m\,(2\,V_{sw})^2/2$ (Bagenal et al. 1997), where $m$ is the mass of the ion and $V_{sw}$ is the solar wind velocity (approximately 450 km $s^{-1}$). Thus, $N^+$ will have energies near 35 keV, and $N_2^+$ will have energies near 70 keV. These are within PEPSSI's energy range, which can measure particles with energies up to 1 MeV. Since the bulk of the pickup ions originate from Pluto, PEPSSI supports the Group 1 objective of measuring the atmospheric escape rate by measuring the number and type of these ions. In addition, the energy range of PEPSSI supports the Group 2 objective of characterizing Pluto's solar wind interaction.

### 3.1.6 SWAP

SWAP is summarized in Weaver et al. and described in detail in McComas et al., this issue. The primary purpose of SWAP is the Group 3 objective of measuring Pluto's interaction with the solar wind. At Pluto, the unperturbed solar wind is expected to have a small proton temperature (i.e., a narrow energy spread), with typical solar wind energies of ~1000 eV. Within the region where the solar wind interacts with Pluto, the energy distribution of the solar wind particles will become less energetic with a wider spread. SWAP's energy range (35 eV to 7.5 keV for the centers of SWAP's electrostatic analyzer bins, McComas et al., this issue) will measure both the unperturbed solar wind, and the solar wind where it is deflected and decelerated due to mass loading by ionized atmospheric gases. Because the spatial scale of the interaction region is directly proportional to the atmospheric escape rate (Bagenal et al. 1997), SWAP supports the Group 1 objective of measuring Pluto's atmospheric escape rate. With an energy range extending to 7.5 keV, SWAP will also detect low energy pick-up ions, overlapping the energy range of the PEPSSI instrument. This allows cross-calibration of the two instruments, and supports the Group 3 objective of measuring Pluto's energetic particle environment.

### 3.1.7 SDC

The Venetia Burney Student Dust Counter is summarized in Weaver et al. and described in detail in Horanyi et al., this issue. The primary purpose of SDC is for Education and Public Outreach, allowing students the opportunity to design, build, fly, operate, and interpret results from a spacecraft instrument. SDC is designed to be the first instrument to measure the dust distribution in our solar system beyond 18 AU, where the dust may show the signature of resonances with Neptune (Fig. 8). Dust measurements are not specified in the science objectives, but are similar to the Group 3 goal of searching for rings, as they relate to Pluto's environment.

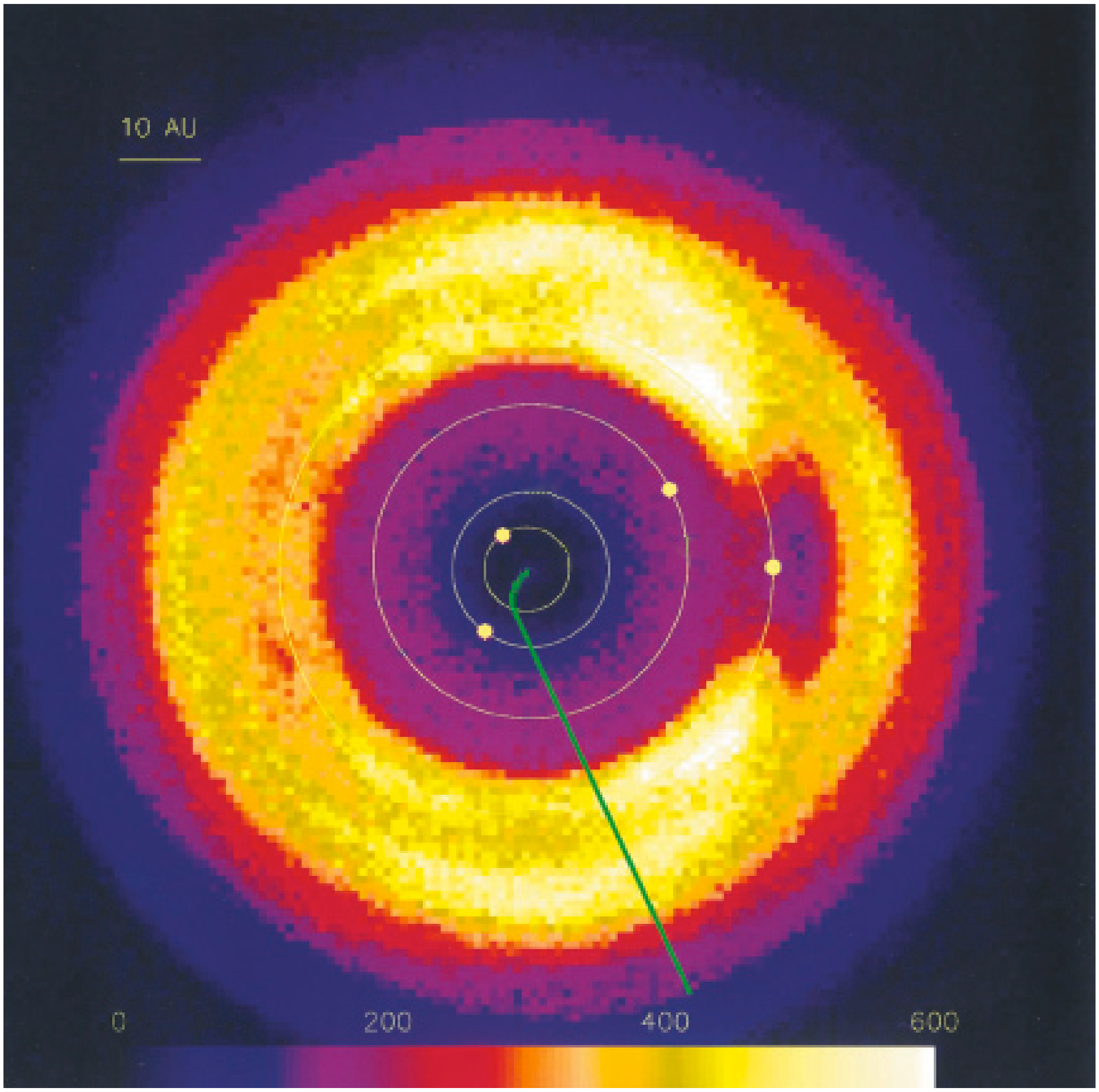

*Figure 8*. Numerical simulation of the column density of 23 μm dust of Liou and Zook (1999, adapted from their Fig 4a), with locations of Jupiter, Saturn, Uranus and Neptune plotted for 2015. The location of New Horizons relative to Neptune is plotted in green.

## 3.2 Mission Design

The trajectory of the New Horizons spacecraft is designed to satisfy the science goals listed in Section 2. Mission design is covered in more detail in Gao and Farquar, this issue. The mission design satisfies the following requirements:

1. New Horizons will arrive at the Pluto system on 2015 July 14. This satisfies the AO-specified requirement that New Horizons arrive at the Pluto system as early as feasible, but before 2020. The arrival time is driven by the change in the sub-solar latitude on Pluto and Charon (and the corresponding decrease in the area on Pluto and Charon illuminated by sunlight), and by the increase in heliocentric distance (which increases the probability of a collapse of Pluto's atmosphere in response to colder ice temperatures).
2. New Horizons will achieve both a Sun and Earth occultation of Pluto. Both occultations are required to meet the Group 1 atmospheric goals. The time of day will allow simultaneous coverage by two NASA deep space network (DSN) stations at elevations >15 deg. Simultaneous coverage by two ground stations will enhance the radio science experiment and lower mission risk (due to weather, for example).
3. New Horizons will achieve a solar occultation by Charon, and, if possible, also an Earth occultation. The Charon solar occultation is required to meet Group 2 Charon atmospheric science objectives.
4. Encounter will occur near opposition, with the Earth between the Sun and New Horizons, at a Sun-Earth-spacecraft angle of 171 degrees. This is greater than the minimum required angle of 45 degrees. An opposition encounter will reduce solar interference with the radio science experiment and decrease the noise floor for the radio occultations.
5. At the time of encounter, Charon will be further from the Sun than Pluto. In this configuration, the nightside of Pluto will be illuminated by reflected sunlight from Charon. This "Pluto-first" encounter timing will enable imaging of the nightside of Pluto.
6. The closest approach distance to the surface of Pluto will be about 10,000 km, which is the encounter geometry for which the remote sensing instruments are designed to achieve their resolution objectives.

## 3.3 Encounter with Asteroid 2002 $JF_{56}$

On 2006 June 13, New Horizons flew past the asteroid (132524) 2002 $JF_{56}$ at a distance of 102,000 km (Olkin et al. 2006a). At the time of closest approach, the asteroid was moving relative to New Horizons at 54 arcsec $sec^{-1}$. This provided an excellent opportunity to test methods for tracking fast-moving objects such as Pluto, since the asteroid had the highest relative rate of any body that New Horizons could image before the encounter with the Pluto system. At the time of encounter, LORRI's door was not yet opened. We therefore carried out observations of the asteroid using both MVIC and LEISA on Ralph. Panchromatic observations were carried out at 35, 13 and 8 hours before closest approach and four-color filter data was collected at 60, 20 and 8 minutes before closest approach. These data include phase angles not achievable from the ground (spanning 4° to almost 90°). Because of the small size of this asteroid and the distant nature of this untargeted flyby, the asteroid was not resolved during any of the observations (at closest approach, an MVIC pixel subtends 2 km at the asteroid), but an unusual (non-stellar) pattern to the signal from the asteroid is consistent with either an irregularly shaped asteroid or a satellite of the asteroid.

## 3.4 Encounter with Jupiter

On 2007 Feb 28, the New Horizons spacecraft will fly past Jupiter at a range of 2.5 million km, and will conduct an extensive series of observations of the planet and its satellites. Highlights of the current observation plan include:

- Time-resolved near-IR image cubes of the Great Red Spot and its surroundings, and high-resolution CCD imaging of the "Little Red Spot"
- Simultaneous UV and near-IR imaging of the Jovian aurorae and airglow
- UV stellar occultations observations of the atmospheres of Jupiter and all the Galilean satellites
- Global imaging of Io's plumes and post-Galileo surface changes
- Global imaging of high-temperature (0.4 - 2.5 micron) volcanic thermal emission from Io
- Observations of UV, and (in eclipse) visible and near-IR, atmospheric emissions from Io, Europa, and Ganymede
- Near terminator imaging of large-scale topographic features on Europa
- 1.25 - 2.5 micron spectroscopy of Galilean satellite surface composition, and similar spatial resolution and better spectral resolution than Galileo NIMS global observations
- Extensive imaging of Jupiter's rings to search for embedded satellites and improve knowledge of the ring structure and ring particle phase functions
- Distant imaging of Himalia and Elara to determine shapes, sizes, and phase curves.
- Plasma observations of the magnetosphere, including an unprecedented 3-month flight down the magnetotail.

## 3.5 Pluto

Plans for the Pluto system encounter are, at this time (some 8 years before the encounter begins) still only preliminary. Nonetheless, the broad outline of our encounter planning is complete. Science observations of Pluto will be divided into five phases. The first phase, the *cruise phase*, includes all observations of the Pluto system in the first nine years of flight. The second phase, the *approach observatory phase*, begins when LORRI can first resolve Pluto, in January 2015. The transition between this and the third phase, the *approach far encounter phase,* is defined by that time when downlink can no longer keep pace with the data collection. For the purposes of this paper, we describe these two phases together as the *approach phase*. Most of the group 1, 2 and 3 objectives are best satisfied in the *near encounter phase*, the period extending from 13 hours before to 5 hours after closest approach. After closest approach there will be a short *departure far encounter phase* of high phase angle observations, radio tracking, and in situ environmental measurements.

### *3.5.1 Cruise phase*

New Horizons observed Pluto with LORRI during instrument commissioning in 2006, and will continue to observe Pluto throughout the cruise period. By 2013, LORRI will be also able to resolve Charon. These observations are of particular benefit to the Group 1 goal of determining the photometric phase functions of Pluto and Charon, since the solar phase angle of Pluto as seen from New Horizons changes over this period, and since New Horizons and Earth observe Pluto from different angles. These observations will also address the Group 2 objective of time variability. As opportunity and spacecraft resources allow, New Horizons will accomplish other science observations as well. For example, New Horizons will be able to observe Uranus and Neptune at phase angles not achievable from Earth-based or near-Earth observatories (Fig. 9).

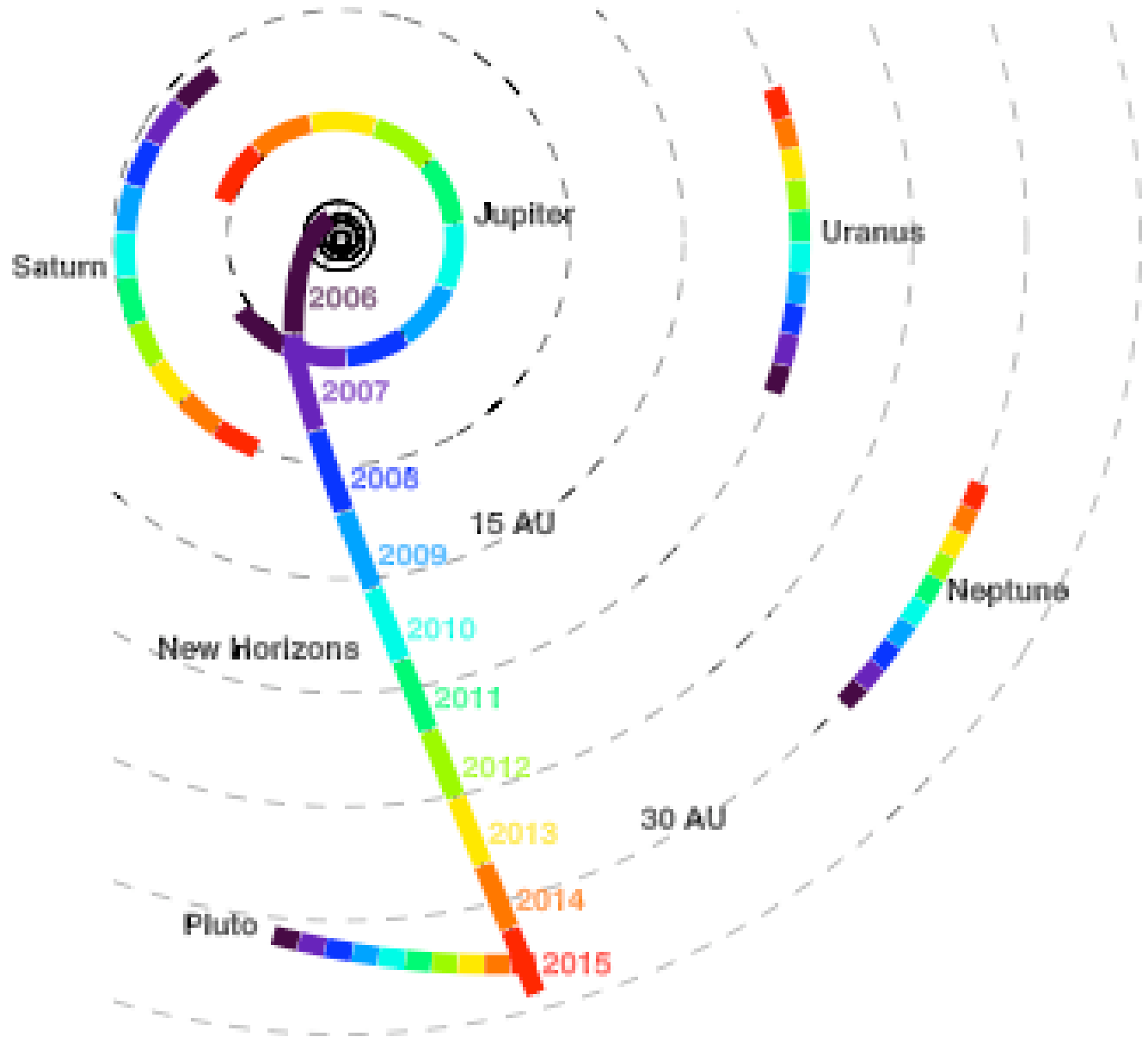

*Figure 9*. Path of New Horizons and the outer planets from launch (2006) to Pluto encounter (2015).

### *3.5.2 Approach phase*

During the approach phase, we will investigate Pluto, Charon, Nix, and Hydra over a long timespan. We will begin imaging the system with MVIC, LEISA, LORRI, and Alice 150 days before closest approach, when LORRI can first resolve Pluto. At this distance, LORRI's resolution will be 450 km pixel$^{-1}$, similar to the to HST maps, MVIC and LEISA will separate Pluto and Charon, and LORRI will easily be able to detect Nix and Hydra and will begin phase curve studies on these satellites. We plan to observe Pluto and Charon every 12 hours from P-150 to P-144 days, covering a complete 6.4-day Pluto rotation period at 30-degree longitudinal spacing. This 6.5-day sequence will be repeated each time the resolution improves by 50%. The current plan is to take eight such sequences, each starting at 150, 100, 66, 44, 28, 19, 12, and 6 days before Pluto closest approach. These observations will address time variability, provide airglow spectra, obtain a series of maps at increasingly higher resolution maps, and refine the orbits and (hence indirectly refine the masses) of Pluto, Charon, Nix, and Hydra. We will also search the returned imagery for small satellites that had previously escaped detection from Earth. The final set of observations, starting at 6 days out, will give the highest spatial resolution images and spectra of the portions of the surface away from the closest-approach hemisphere (including the far-side hemisphere at 3.2 days before closest approach). One or two 27-day solar rotations before closest approach, SWAP will begin solar wind observations to allow interpolation of the unaffected solar wind upstream of Pluto. The PEPSSI campaign to detect pick-up ions from Pluto's atmosphere will begin during the final days before closest approach, possibly detecting Pluto ions as far as 1 million km upstream of Pluto.

### *3.5.3 Near encounter phase*

The strawman near encounter phase was designed to fulfill the objectives outlined in Table III. While encounter planning will be readdressed prior to the Pluto encounter rehearsals in 2012, a detailed plan was formed pre-launch in 2001-2003, to guide spacecraft design, instrument placement, pre-launch tests, commissioning, and checkout. What we present below (Table V) is based on that strawman sequence, updated for expected spacecraft performance (e.g., pointing accuracy, navigation error ellipses, and settle times) in October 2005.

Near the start of the near encounter phase, roughly 13 hours before Pluto closest approach, LORRI will take images of Pluto and Charon where each body just fits within a single 1024x1024 pixel frame, giving high geometric fidelity of the entire approach hemisphere at 2-3 km pixel$^{-1}$. Much of the next seven hours will be spent taking approach airglow spectra with Alice, mainly for the Group 1 goal of atmospheric CO and Ar detection. Additional LORRI mosaics will be taken before LEISA takes the first set of Group 1 surface composition maps at about 3 hours before closest approach, at 10 km pixel$^{-1}$. Additional Alice observations will follow; Alice will resolve the surfaces of Pluto and Charon at this time. At two to 1.5 hours before closest approach, LEISA will take a second, redundant set of Group 1 surface composition maps at 5-7 km pixel$^{-1}$. New Horizons will then obtain the Group 1 geology (panchromatic and color) maps, with high-resolution visible and infrared images taken near Pluto and Charon closest approaches.

After closest approach, New Horizons will pass through the occultation zones of Pluto and Charon. Enough time is allocated for Alice to establish an unocculted solar spectrum and for REX to establish an upper baseline before and after Pluto's ionosphere is detectable. REX and Alice will observe the Earth and Sun occultations simultaneously. During the solid body occultation, REX will observe the thermal emission from Pluto or Charon. Immediately before and after the two occultations, high-phase angle images by MVIC will be used to search for hazes and rings in forward scattering, as well as attempt imaging of Pluto's night side in reflected Charon light. Although the particle instruments usually ride along with the pointing determined by the remote sensing instruments, a roll is planned between the two occultations specifically for SWAP and PEPSSI, when New Horizons will be downstream of Pluto and Charon. (The term "Pam" in the instrument description refers to both PEPSSI and SWAP; Stern and Cheng 2002). Also after closest approach, Alice will look for nightglow emission from Pluto, and LESIA will take high phase angle observations of both Pluto and Charon.

TABLE V: Strawman near encounter sequence

| Start Time re Pluto CA (min) | Target | Instrument | Description | Highest Group (1,2,3) | Start solar phase angle (deg) | Starting Resolution (km) | Start range to target center (km) |
|---|---|---|---|---|---|---|---|
| -800.0 | Pluto | LORRI | P_pan_global_1x1_last | 3 | 16.0 | 3.30 | 661,285 |
| -589.7 | Pluto | ALICE | PC_UV_airglow_spectrum_1 | 1 | 16.3 | 2431.5 | 487,484 |
| -487.9 | Charon | LORRI | C_pan_global_1x1_last | 3 | 18.2 | 2.09 | 417,749 |
| -486.1 | Pluto | ALICE | PC_UV_airglow_spectrum_2 | 1 | 16.6 | 2003.6 | 401,902 |
| -385.2 | Pluto | ALICE | PC_UV_airglow_spectrum_3 | 1 | 17.0 | 1586.8 | 318,536 |
| -284.2 | Pluto | ALICE | PC_UV_airglow_spectrum_4 | 1 | 17.7 | 1169.8 | 235,131 |
| -243.2 | Pluto | LORRI | P_pan_global_3x3_1 | 2 | 18.2 | 1.00 | 201,289 |
| -236.1 | Pluto | ALICE | PC_UV_airglow_spectrum_5 | 1 | 18.2 | 971.3 | 195,444 |
| -215.1 | Pluto | LORRI | P_pan_global_3x3_2 | 2 | 18.6 | 0.88 | 178,118 |
| -206.1 | Pluto | LEISA | P_IR_global_scan_1 | 1 | 18.7 | 10.5 | 170,695 |
| -198.6 | Pluto | LEISA | P_IR_global_scan_2 | 1 | 18.9 | 10.1 | 164,510 |
| -191.0 | Pluto | LEISA | P_IR_global_scan_3 | 1 | 19.0 | 9.7 | 158,257 |
| -183.3 | Charon | LEISA | C_IR_global_scan_2 | 1 | 23.9 | 10.3 | 166,060 |
| -177.2 | Pluto | ALICE | PC_UV_airglow_spectrum_6 | 1 | 19.3 | 728.4 | 146,854 |
| -165.5 | Pluto | ALICE | P_UV_surface_spectrum_1 | 2 | 19.6 | 680.3 | 137,241 |
| -153.7 | Charon | ALICE | C_UV_surface_spectrum_1 | 2 | 25.5 | 706.2 | 141,839 |
| -140.7 | Pluto | LEISA | P_IR_global_doublescan_1 | 1 | 20.4 | 7.2 | 116,796 |
| -131.8 | Pluto | LEISA | P_IR_global_doublescan_2 | 1 | 20.8 | 6.7 | 109,530 |
| -122.8 | Pluto | LEISA | P_IR_global_doublescan_3 | 1 | 21.2 | 6.3 | 102,064 |
| -113.3 | Pluto | LEISA | P_IR_global_doublescan_4 | 1 | 21.8 | 5.8 | 94,263 |
| -104.3 | Charon | LORRI | C_pan_global_6x3_1 | 2 | 29.8 | 0.51 | 101,837 |
| -88.7 | Charon | LEISA | C_IR_global_doublescan_1 | 1 | 32.0 | 5.5 | 89,326 |
| -78.0 | Charon | LEISA | C_IR_global_doublescan_2 | 1 | 33.9 | 5.0 | 80,901 |
| -66.3 | Charon | MVIC | C_color_global_scan_1 | 1 | 36.5 | 1.42 | 71,809 |
| -62.3 | Charon | MVIC | C_color_global_scan_2 | 1 | 37.5 | 1.36 | 68,667 |
| -59.1 | Pluto | LORRI | P_pan_alongtrack_8x2_2 | 1 | 27.8 | 0.24 | 50,082 |
| -44.8 | Pluto | MVIC | P_color_global_scan_1 | 1 | 31.7 | 0.75 | 38,619 |
| -39.6 | Pluto | MVIC | P_color_global_scan_2 | 1 | 33.7 | 0.67 | 34,550 |
| -34.1 | Pluto | MVIC | P_pan_global_scan_1 | 1 | 36.5 | 0.58 | 30,281 |
| -29.0 | Pluto | MVIC | P_pan_global_scan_2 | 1 | 39.9 | 0.50 | 26,417 |
| -23.5 | Pluto | LEISA | P_IR_hires_scan_1 | 2 | 44.7 | 1.3 | 22,372 |
| -14.2 | Pluto | LORRI<br>MVIC | P_pan_terminator_simultMVIC_1 | 2 | 58.5 | 0.07<br>0.30 | 16,139 |
| -4.0 | Charon | LEISA | C_IR_hires_scan_1 | 2 | 75.7 | 1.9 | 30,517 |
| 4.5 | Charon | LORRI<br>MVIC | C_pan_terminator_simultMVIC_1 | 2 | 88.6 | 0.14<br>0.54 | 27,785 |
| 13.1 | Charon | MVIC | C_pan_global_scan_1 | 1 | 103.5 | 0.52 | 26,693 |
| 20.1 | Pluto | MVIC | P_pan_GlobalHaze_scan_1 | 1 | 161.3 | 0.38 | 19,969 |
| 25.8 | Pluto | ALICE | P_UV_occ_solar_1 | 1 | 167.6 | 7.6 | 24,067 |
| 25.8 | Pluto | REX | P_radio_occ_simultALICE_1 | 1 | 167.6 | 1.0 | 24,067 |
| 76.1 | Pluto | REX | P_radio_temperature_map_1 | 2 | 175.0 | 1338.5 | 63,860 |
| 83.3 | Pluto | PEPSSI+SWAP | P_particles_roll_simultPAM_1 | 3 | 174.1 | | 69,769 |
| 117.9 | Pluto | ALICE | P_UV_nightglow_spectrum_2 | 2 | 171.4 | 484.6 | 98,094 |
| 129.8 | Charon | ALICE | C_UV_occ_solar_1 | 2 | 179.3 | 31.3 | 100,262 |
| 129.8 | Charon | REX | C_radio_occ_simultALICE_1 | 2 | 179.3 | 2.1 | 100,262 |
| 148.0 | Pluto | MVIC | P_pan_GlobalRings_scan_1 | 3 | 170.1 | 2.43 | 122,831 |
| 155.0 | Pluto | ALICE | P_UV_nightglow_spectrum_3 | 2 | 169.9 | 637.1 | 128,593 |
| 282.0 | Pluto | LEISA | P_IR_hiphase_scan_1 | 2 | 167.7 | 14.4 | 233,342 |
| 292.0 | Charon | LEISA | C_IR_hiphase_scan_1 | 2 | 171.8 | 14.4 | 233,328 |

Blue=Alice, Green = MVIC or LORRI, Red = LEISA, Gold = REX, Black = PEPSSI or SWAP. Grey shading = Charon. SWAP and PEPSSI on throughout the near encounter as power allows.

#### *3.5.4 Departure phase*

The departure phase is currently planned to extend for 30 days past Pluto closest approach, approximately four Pluto rotations. This period will allow SWAP and PEPSSI to measure in situ particle populations for one solar rotation downstream of Pluto. In departure, SDC will search for impact rate enhancements associated with dust from the Pluto system. REX will measure globally averaged temperatures for some days after closest approach. The strawman plan also calls for a series of remote sensing observations at the 165 degrees phase angle asymptote during departure over one Pluto rotation.

## 3.6 Kuiper Belt Objects.

The New Horizons mission is designed to make flybys of one or more KBOs after Pluto. The particular Kuiper Belt Objects for New Horizons are not yet selected. Thanks to a favorable launch, we expect to have approximately 200 m $s^{-1}$ available to target one or more KBO flybys. With this amount of fuel, there is a cone of accessibility with a full-width of 1.6 deg around New Horizon's nominal trajectory. In this volume, there should be a 50% chance of reaching a KBO 72 km or larger, or 95% of reaching one 45 km or larger (Spencer et al. 2003). The probability for a KBO encounter as a function of distance along the New Horizons path is proportional to the number of KBOs per cubic AU and the width of the cone. This probability is strongly peaked at 42 AU. It is reasonable to expect that a post-Pluto KBO flyby to occur in 2018 or 2019.

The planned observations at KBOs are similar to those at Pluto. With MVIC, LORRI, and LEISA we can obtain visible color and panchromatic maps and IR composition maps of each flyby target, yielding shape, geological/geomorphological, photometric, cratering history, and compositional data, and we will search for evidence of volatiles, a past or present atmosphere, and potential satellites. In addition, LORRI can obtain long-range images to search for small satellites and globally map the object during approach, and high-resolution imagery at closest approach in order to study surface cratering and regolith processes, volatile transport, etc., at spatial scales down to 100 m or less (depending on the closest approach distance). Regarding any potential atmosphere, Alice can obtain airglow data like that obtained at Pluto and SWAP and PEPSSI can search for evidence of a solar wind interaction zone and upstream pickup ions (perhaps even sputtered from the surface). We also expect to have sufficient targeting accuracy to search for tenuous extended atmospheres using Alice for solar occultations, considering that any such atmosphere should be loosely bound. Doppler tracking of New Horizons will measure or place limits on the KBO mass. REX will measure brightness temperatures at 4.2 cm. Finally, SDC will measure the dust in the outer solar system, providing a different approach to understanding the collisional evolution of the Kuiper Belt.

# 4. Meeting the Mission Scientific Objectives at the Pluto System

The proposed mission trajectory, instruments, and strawman observing sequence fulfill the Group 1, 2 and 3 objectives listed in Table III, as summarized in Table VI.

Table VI
Mapping of science objectives to instruments

| Objective | | Measurement technique: Primary | Measurement technique: Supporting |
|---|---|---|---|
| **Group 1** | | | |
| Geology/ Geophysics | pan | MVIC, LORRI | LEISA |
| | color | MVIC | LEISA |
| Surface composition | | LEISA | MVIC, Alice |
| Neutral atmosphere | | Alice, REX | MVIC, LORRI, PEPSSI, SWAP |
| **Group 2** | | | |
| Variability | | MVIC, LEISA, Alice, LORRI, REX | PEPSSI, SWAP |
| Stereo | | MVIC, LORRI | LEISA |
| Hi-res terminator maps | | MVIC, LORRI | LEISA |
| Hi-res composition maps | | LEISA | MVIC |
| Ionosphere/solar wind | | REX, SWAP | PEPSSI |
| Other atmospheric species | | Alice | PEPSSI, SWAP |
| Charon atmosphere | | Alice | REX, LEISA |
| Bond albedos | | MVIC, LORRI | LEISA |
| Surface temperatures | | LEISA | REX |
| **Group 3** | | | |
| Energetic particles | | PEPSSI | SWAP |
| Bulk parameters | | REX, LORRI | MVIC, LEISA |
| Magnetic field search | | not directly addressed | SWAP |
| Satellite and ring search | | MVIC, LORRI | Alice, REX, SDC |

## 4.1 Group 1: Objectives required for mission success

### *4.1.1 Characterize the global geology and morphology of Pluto and Charon*

**Panchromatic mapping.** Every panchromatic observation will be folded into the global maps, including images of Pluto and Charon, beginning one full rotation period (6.4 days) before closest approach, when LORRI can resolve Pluto and its satellites at 36 km per pixel. We plan to take the main Group 1, 0.5 km pixel$^{-1}$ maps of Pluto and Charon redundantly. For Charon, LORRI will make one mosaic map at low phase (C_pan_global_6x3_1) and MVIC will make another with by scanning the pan TDI array near Charon closest approach (C_pan_global_scan_1). MVIC will take the Group 1 0.5-km pixel$^{-1}$ map of Pluto with two redundant panchromatic TDI scans about 30 minutes before closest approach (P_pan_global_scan_1-2). In the current plan, LORRI will augment MVIC's 0.5 km pixel$^{-1}$ hemispheric maps of Pluto with hemispheric mosaics at 1.0 and 0.9 km pixel$^{-1}$ (P_pan_global_3x3_1-2), and a 400-km wide swath across the nadir of Pluto at 0.24 km pixel$^{-1}$ (P_pan_alongtrack_8x8_2). Higher resolution hemispheric LORRI maps may be possible, depending on spacecraft pointing. As shown in Reuter et al. and Cheng et al (this issue), the signal-to-noise will meet the AO requirement of 50 or higher per pixel.

**Color mapping.** MVIC will make redundant four-color images of the Pluto system with color TDI scans. Pluto global color maps are taken at ~0.7 km pixel$^{-1}$ (P_color_global_scan_1-2), while Charon global color maps are taken at 1.4 km pixel$^{-1}$ (C_color_global_scan_1-2), exceeding the AO spatial resolution requirement of 1.5-5 km pixel$^{-1}$. As shown in Reuter et al. (this issue), the signal-to-noise will meet the AO requirement of 50 or higher per pixel in two or more of the color channels.

**Phase Angle Coverage.** The phase angles to Pluto and its satellites during the approach phase will increase slowly from 14.3 to 15 degrees, then increase quickly during the near encounter phase, as the spacecraft swings around Pluto and Charon. Phase angles for visible imaging before or near closest approach will be between 16 and 60 deg for Pluto and between 18 and 104 deg for Charon. Currently, we have observations of Pluto planned shortly after closest approach, with a phase angle near 160 deg, designed mainly for haze detection (see Section 4.1.3). The capability of LORRI and Ralph to image at this phase angle, and the even higher 165 deg departure asymptote, has not yet been confirmed by in-flight tests.

### *4.1.2 Map surface composition of Pluto and Charon*

Ralph/LEISA will produce the main global spectral maps of Pluto and its satellites, producing spectral data cubes at a spectral resolution of ~250 in the interval 1.25-2.5 micron, and at ~560 in the interval 2.10-2.25 micron. LEISA will take hemispheric spectral images of both Pluto and Charon at 10 km pixel$^{-1}$ about 3 hours before closest approach (P_IR_global_scan_1-3, C_IR_global_scan_2). At about 2 hours before closest approach, a redundant set of spectral images for Pluto will be made at resolutions of 5-7 km pixel$^{-1}$ (P_IR_global_doublescan_1-P_IR_global_doublescan_4), and at about 1 1/2 hour before closest approach, the redundant hemispheric Charon observation are made at resolutions of 5 km pixel$^{-1}$ (C_IR_global_doublescan_1-2). All spectral data cubes exceed the AO spatial resolution requirement of 5-10 km pixel$^{-1}$. Observations of Nix and Hydra remain will be planned after better orbit solutions are obtained. For each global map, LEISA scans over the target up to four times, both to increase the signal-to-noise ratio and to cover the entire target when it subtends more than 256 pixels. These hemispheric observations are made at phase angles at 15-35 deg. The hemispheric spectral images are augmented by regional spectral images near closest approach, at resolutions of 1-2 km pixel$^{-1}$ and phase angles of 45 deg (Pluto) and 75 deg (Charon). If feasible, observations will be taken at high phase angle on departure, at resolutions near 14 km pixel$^{-1}$. At the expected SNR for the planned spectra (Reuter et al., this issue), LEISA will map the spatial distribution of $N_2$, CO, $CH_4$, and other species (Table VII).

TABLE VII
Ralph/LEISA pre-flight estimated IR detection limits for surface ices

| Molecule | 3-σ limit | Comments |
|---|---|---|
| $N_2$ (2.15 μm) | 20% | Pure $N_2$ at 5x5 binning |
| $N_2$ detected by $CH_4$ band shifts | 5% | For CH4 abundance <2% |
| $CH_4$ (pure or diluted 2.2 μm) | 0.1% | Detect either mixing state |
| $CH_4$ vs. $CH_4$:$N_2$ (e.g., 2.2 μm) | 10% | Distinguish mixing states |
| $H_2O$, crystalline (e.g., 1.65 μm) | 10% | Quite model dependent |
| $H_2O$, amorphous (e.g., 2.0 μm) | 10% | Quite model dependent |
| $NH_3$ & hydrates (2.20–2.25 μm) | 5% | Quite model dependent |
| CO (2.35 μm) | 0.1% | Probably dissolved in N2 |
| $CO_2$ (1.96 μm) | 2% | Probably not in N2 |
| $CH_3OH$ (2.28 μm) | 25% | Better if dissolved in N2 |
| $C_2H_6$, $C_2H_6$:$N_2$ (1.68, 2.33 μm) | 0.1% | Better if dissolved in N2 |
| $C_2H_2$, $C_2H_2$:$N_2$ (2.45, 2.42 μm) | 0.1% | Better if dissolved in N2 |
| $C_2H_4$, $C_2H_4$:$N_2$ (2.22, 2.26 μm) | 0.1% | Better if dissolved in N2 |

Limits are conservative; we assumed 1x1 binning unless noted, intimate mixing with other expected species, and realistic albedos. We will also search for $SO_2$ (2.13 μm), $H_2S$ (1.64 μm), $H_2CO$:$N_2$ (2.18 μm), HCN (1.91 μm), $HC_3N$ (1.83 μm), pyroxene (1.79–2.33 μm), & kaolinite-serpentine clays (1.40 μm).

### *4.1.3 Characterize the neutral atmosphere of Pluto and its escape rate.*

**Compositional Measurements:** Alice is the instrument New Horizons will use for measuring the neutral atmospheric composition at Pluto. $N_2$ and $CH_4$ will be measured through the solar UV occultation (P_UV_occ_solar_1). Stellar occultations are also being planned, but have not yet been identified. These occultations will provide altitude-resolved abundance measurements at both the ingress and egress sites. $N_2$ densities are measured through continuum absorption and electronic transitions from 465-1000 Å. $CH_4$ will be detected by its continuum absorption at 1000-1300 Å. Airglow observations with Alice on approach (PC_UV_airglow_spectrum_1-6) will be the main technique for measuring CO and Ar, although both species may be detectable in the occultation spectra as well. The Fourth Positive Group (14,4) airglow band at 1356 Å, excited by resonance fluorescence of the (14,0) band by solar Lyα, should be the brightest CO feature above the level where $CH_4$ absorbs FUV. Ar will be searched for at its well-known EUV resonance lines (1048 and 1067Å).

**Upper atmospheric thermal structure.** The primary data set for the Group 1 objective of measuring the thermal and pressure structure in Pluto's upper atmosphere comes from Alice solar and stellar occultations (P_UV_occ_solar_1). Since $N_2$ is expected to be the dominant species, we will derive temperatures from the line-of-sight number density of $N_2$ under the assumption of hydrostatic equilibrium. From our simulations (based on the M2 model atmosphere of Krasnopolsky and Cruikshank, 1999), we will be able measure temperatures and thermal gradients to the AO-required accuracy (10% in 100 km bins) from altitudes of 1300 to 1800 km using only the $N_2$ continuum absorption shortward of 665 Å. By using the $N_2$ lines in addition to the $N_2$ continuum absorption, we will extend this range downward for several hundred km, until altitudes where $CH_4$ absorption becomes important.

**Lower atmospheric thermal structure.** Pluto's atmospheric pressure, number density and temperature near the surface will be measured with radio occultations (P_radio_occ_simultALICE_1). REX radio occultation profiles will measure atmospheric phase delays to better than 1 deg (or 0.05 of the estimated phase delay at 3 microbar). Pluto's atmospheric pressure at the time of arrival is uncertain, but the recent optical occultations of 2002 (Elliot 2003, Sicardy 2003) and 2006 (Young 2007, Elliot et al 2007) suggest that a 4 microbar surface pressure is a conservative estimate. For this case, pre-flight models estimate that REX will measure the base number density ($n$) to 1%, temperature ($T$) to 1.5 K, and pressure ($p$) to 0.2 microbar at 12 km vertical resolution (less than half a scale height) in a single occultation with only one REX operating and one DSN uplink. By using a second DSN station and combining ingress and egress occultations, REX will measure $n$ to 0.5%, $T$ to 0.75 K, and $p$ to 0.1 μbar. See Tyler et al. (this issue) for updated signal-to-noise estimates.

**Evolution and escape rate.** Our objective here is to quantify Pluto's atmospheric loss rate (currently uncertain to over an order of magnitude; McNutt 1989; Trafton et al. 1997. Bagenal et al. 1997; Krasnopolsky 1999; Tian and Toon 2005), and determine whether the atmosphere is hydrodynamically escaping. Critical parameters to measure are the atmospheric scale heights of the escaping species and the total 500-1850 Å solar energy input that heats Pluto's thermosphere. Escape calculations suggest that Alice occultation profiles obtained by measuring these quantities upward from near the energy deposition peak (i.e., $\sim 3 \times 10^8$ $cm^{-3}$ to $\sim 10^7$ $cm^{-3}$, perhaps 1300-2300 km altitude) can allow the determination of the species integrated escape flux to ~35%. The scale height of $N_2$ high in Pluto's atmosphere will be measured by solar occultation (P_UV_occ_solar_1). Also, Alice H Ly α airglow scale height profiles (from all Alice airglow measurements) can be used to derive an independent H-only escape flux (Clarke et al. 1992).

If the escape from Pluto's atmosphere is substantial (e.g. greater than about 1 x $10^{27}$ molecules per second) then ionization of the escaping neutrals will cause a measurable perturbation of the solar wind that extends well away from Pluto. The New Horizons plasma instruments will measure both the spatial extend and nature of the solar wind interaction with Pluto's escaping atmosphere. In particular, PEPSSI will measure the flux of Pluto pickup ions and SWAP will identify and measure the location of the atmosphere/solar wind interface (so long as we approach close enough to Pluto). Because pickup protons from ionized interstellar H are ubiquitous in the outer solar system, it is necessary to distinguish them from Pluto's pickup ions; PEPSSI energetic particle sensor has the mass resolution to separate interstellar $H^+$ from heavy $CH_4^+$, $N_2$+, $CO^+$ and other molecular ions escaping from Pluto's atmosphere.

**Aerosols and haze detection.** At Triton, Voyager 2 imaged both a uniform hydrocarbon haze (with a scale height of 10 km and a vertical optical depth of 0.001–0.01), and also patchy condensed $N_2$ clouds at altitudes <10 km with vertical optical depth near 0.1 (Rages and Pollack 1992). Based on Pluto's expected thermal profile, condensation clouds may be unlikely, but a hydrocarbon haze is expected by many. All low phase-angle approach maps of Pluto will be searched for clouds (e.g., P_pan_global_1x1_last, P_pan_global_scan_1). Additionally, the Alice solar occultation (P_UV_occ_solar_1) will achieve 15% accuracy in measuring haze with vertical optical depth of 0.005 in 5-km height bins above 500km, using 1800–1850 Å wavelengths. After closest approach, MVIC images will be obtained at high phase angles to further characterize hazes by their forward scattering properties (P_pan_GlobalHaze_scan_1). Even Pluto's haze is ten times less absorbing than Triton's, MVIC will be able to detect haze from the surface to 80 km above the surface at a phase angle of 135 deg. Higher phase angle measurements will also be attempted.

## 4.2 Group 2 Objectives:

### *4.2.1 Characterize the time variability of Pluto's surface and atmosphere*

For variability on timescales of a Pluto day (6.4-day rotation period), LEISA will use the temperature-sensitive $N_2$, $CH_4$, and $H_2O$ bands to look for surface temperature changes as a function of local time of day and unit type, supported by REX radiometry of disk-averaged brightness temperatures at 4.2 cm on both day and night sides of Pluto. LEISA composition mapping and LORRI and MVIC imaging will be sensitive to differences in albedo and composition of areas at different times of day, perhaps showing the effects of ices condensing overnight. We will also use the combined entry and exit Earth radio occultation datasets (combined with imagery and IR spectroscopy of these locales) to search for variability at two local times of day in both the neutral atmosphere and the ionosphere. Spatially resolved UV airglow spectra of day and night-side atmospheres (and ingress and egress occultation profiles) will also be compared to search for diurnal and spatial variability in the neutral atmosphere.

New Horizons will begin observing Pluto 150 days before closest approach, at which time LORRI resolves Pluto is resolved with a diameter of 2.6 pixels across. Temporally resolved spectra and photometry during the approach phase will be searched for evidence of change. In the final three Pluto rotations, starting at 19 days before closest approach, the optical instruments will be able to track meteorology and cloud motions at rates as slow as 1 m $s^{-1}$ and look for variations in the style or degree of geyser-like activity.

Observations during cruise may detect changes on longer, seasonal timescales. The powerful combination of surface composition maps, visible albedo maps, and atmospheric profiles will constrain seasonal models of Pluto's interacting atmosphere and volatile distribution.

MVIC and LORRI imaging will be used to study variations in surface unit age via stratigraphic relationships and crater density variations, and to search for wind streaks, dunes, or other aeolian features that can serve to suggest or place limits on past epochs of higher atmospheric bulk. Sublimation scarps may indicate how much frost escaped from Pluto over the age of the solar system.

### 4.2.2 Image Pluto and Charon in stereo

Topography from stereo imaging will be inferred from comparison of images obtained from two vantage points, for which topographical height differences between two control points will have a more noticeable affect at larger horizontal separations. Throughout the ten hours before closest approach, large portions of Pluto and Charon are imaged at a range of viewing angles, and all such images will be used for stereo reconstruction. One such pair, P_pan_global_3x3_1 and P_pan_global_3x3_2, will allow height reconstruction of 1 km for features separated by 1000 km, or 5 km for features separated by 250 km. In addition to stereo reconstruction, topography can be deduced from shape-from-shading (photoclinometry), which relies on the photometric phase function (goal 1.1c).

### 4.2.3 Map the terminators of Pluto and Charon with high resolution

At roughly an hour before Pluto closest approach, LORRI will observe a strip crossing the terminator across Pluto roughly 1900 long by 450 km wide, at a phase angle of 28 deg with a resolution of 240 m pixel$^{-1}$ (P_pan_alongtrack_8x2_2). The highest resolution Pluto images will be taken fifteen minutes before Pluto closest approach, when LORRI and MVIC will take a series of simultaneous images of Pluto at 59 degrees phase (P_pan_terminator_simultMVIC_1). Six to ten images will be made of Pluto by each instrument, depending on the size of the error ellipse, which is dominated by the time of flight uncertainty. With LORRI, the resolution will be 0.70 km pixel$^{-1}$, so each image will cover 72 x 72 km. MVIC's resolution at the same distance is 0.30 km pixel$^{-1}$, and each image will cover 38 x 1500 km rectangular areas.

As with Pluto, Charon will also be imaged with LORRI and MVIC for high-resolution terminator images. This is planned to occur an hour and a half before closest approach at a phase angle of 30 deg with LORRI (C_pan_global_6x3_1), and again simultaneously with MVIC and LORRI near closest approach at a phase angle of 89 deg (C_pan_terminator_simultMVIC_1).

### 4.2.4 Map the surface composition of selected areas of Pluto and Charon with high resolution

Near Pluto closest approach, LEISA will take a high spatial resolution scan across Pluto at 1.3 km pixel$^{-1}$, obtaining spectra on a strip measuring 824 km by 332 km (P_IR_hires_scan_1). The similar high spatial resolution LEISA scan of Charon will achieve a resolution of 1.9 km pixel$^{-1}$, over a strip measuring 1033 km by 486 km (C_IR_hires_scan_1). Additionally, MVIC will obtain approach hemisphere color maps of both bodies at a resolution (at the sub-spacecraft point) of 0.7 km pixel$^{-1}$ for Pluto (P_color_globalscan_1-2) and 1.4 km pixel$^{-1}$ for Charon (C_color_globalscan_1-2). These color maps will be obtained in all four MVIC color bands, including the $CH_4$ filter at 0.89 micron. Because the MVIC color maps and the high-resolution LEISA maps have resolutions comparable to the resolutions of the global panchromatic maps obtained as a Group 1 objective, they will allow us to relate compositional information with geology.

### *4.2.5 Characterize Pluto's ionosphere and solar wind interaction*

REX will search for the signature of an ionosphere around Pluto during its Earth occultation (P_radio_occ_simultALICE_1), and will be sensitive to number densities of roughly 2000 $e^-$ $cm^{-3}$. SWAP and PEPSSI also contribute to this goal. Even if the atmospheric escape rate from Pluto is small (<1 x $10^{27}$ molecules $s^{-1}$) the ionosphere forms a conducting barrier to the supersonic solar wind, with a bow shock will forming upstream. The weak interplanetary magnetic field of the outer solar system means that the shock is likely to be very broad so the New Horizons' particle detectors should be able to detect the shock and perhaps measure structure within it. SWAP will also search for wake effects after closest approach (P_particles_roll_simultPAM_1, and during P_UV_occ_solar_1). By comparing low-energy data from SWAP with pickup ions detected by PEPSSI, we will be able to determine outflow rate from the planet's atmosphere.

### *4.2.6 Search for neutral species including H, $H_2$, HCN, and $C_xH_y$, and other hydrocarbons and nitriles in Pluto's upper atmosphere, and obtain isotopic discrimination where possible*

Some minor species, such as H and Ne, will primarily be detected in Alice UV airglow measurements (PC_UV_airglow_spectrum_1-6). Other species will be detected from their opacity profiles in the Alice UV solar occultation data (P_UV_occ_solar_1). We should be able to detect HCN at mixing rations greater than $2x10^{-4}$, $C_2H_2$ at $1x10^{-5}$, $C_2H_4$ at $7x10^{-6}$, and $C_2H_6$ at $3x10^{-4}$.

### *4.2.7 Search for an atmosphere around Charon*

The surface gravity of Charon is less than half that of Pluto's. As a result, its primordial atmosphere is expected to have escaped since Charon's formation, and a secondary atmosphere on Charon from Pluto's escaping atmosphere is also unlikely (Trafton et al. 1997). While analysis of a 1980 stellar occultation by Charon suggested the possibility of a tenuous atmosphere based on two anomalously low points in the upper baseline (Elliot and Young 1991), a stellar occultation in 2005, at much higher time resolution, showed no evidence for an atmosphere (Gulbis et al. 2005, Sicardy et al. 2005, Young et al. 2005), with upper limits near $2x10^{20}$ $cm^{-2}$. Our encounter plan includes Alice UV observations of a solar occultation by Charon (C_UV_occ_solar_1), which will be able to detect an atmosphere down to the nanobar range if it is present. While precedence is given to Alice in the targeting of the Charon occultation, a radio occultation may be observable as well (C_radio_occ_simulALICE_1).

### *4.2.8 Determine bolometric Bond albedos for Pluto and Charon*

The bond albedo is the ratio of sunlight reflected in all directions to incident sunlight; it is relevant for calculating energy balance at the surface. The strict definition of bond albedo is a global quantity, derived from the disk-integrated brightness observed at a range of solar phase angles. Since Pluto is inhomogeneous in albedo, composition, and temperature, energy balance depends on the local version of bond albedo, namely normal reflectance and photometric phase functions. These rely on disk-resolved observations at a range of phase functions.

During the approach phase, New Horizons will make UV, visible, and infrared observations of Pluto, Charon, Nix, and Hydra at phase functions ranging from 14.1 to 14.9 deg. In the final week prior to closest approach, LORRI will observe Pluto, Charon, Nix and Hydra every 12 hours at resolutions better than 36 km. In the near encounter sequence, New Horizons will observe the full visible disks of Pluto and Charon with resolution better than 3 km $pixel^{-1}$ at phase angles ranging from 16 to 58 deg (Pluto) and 16 to 104 deg (Charon). Higher phase observations are scheduled in the strawman sequence for phase angles ranging from 160 to 170 deg.

#### *4.2.9 Map the surface temperatures of Pluto and Charon.*

We will measure the temperatures of Pluto, Charon, Nix, and Hydra using LEISA to observe IR spectral features that are diagnostic of temperature in all LEISA spectral images, at a resolution up to 1.3 km $pixel^{-1}$ (P_IR_hires_scan_1). REX will measure the brightness temperature at 4.2 cm on a much coarser resolution scale (P_radio_temperature_map_1, but also passive radiometry during the solid body occultations of P_radio_occ_simultALICE_1 and C_radio_occ_simultALICE_1).

### 4.3 Group 3 Objectives

#### *4.3.1 Characterize the energetic particle environment of Pluto and Charon*

For large atmospheric escape rates, the interaction may be best described as “comet-like,” with significant mass-loading over an extensive region; for small escape rates the interaction is probably confined to a much smaller region, creating a more “Venus-like” interaction (Luhmann et al., 1991), where electrical currents in the gravitationally bound ionosphere deflect the solar wind flow. At aphelion (50 AU), should Pluto’s atmosphere completely collapse and freeze onto the surface, then the interaction becomes “Moon-like” with the solar wind suffering minimal deflection and directly bombarding the bare, icy dayside surface. Not having a detectable atmosphere, Charon almost certainly has such a “Moon-like” interaction, remaining primarily in the solar wind if Pluto’s interaction is weak but becoming totally engulfed if Pluto’s interaction is strongly “comet-like” and extends beyond Charon’s orbit at 17 $R_P$ (where Pluto’s radius, $R_P$, is 1150 km). The solar wind is supersonic so that when the flow impinges on an obstacle (such as the magnetosphere of the Earth or other planet) an upstream bow shock must form to slow and deflect the supersonic (actually superfast-mode magnetosonic) plasma. The weak interplanetary magnetic field (IMF) at 30 AU and heavy ions formed by photoionizing the heavy molecules of Pluto’s escaping atmosphere have very large gyroradi (~500 $R_P$). The net results of these non-fluid or kinetic effects are to make the bow shock a thick transition region and to make the shape of the interaction region asymmetric where the direction of asymmetry is governed by the direction of the IMF. The New Horizons’ SWAP instrument will measure the perturbation of the solar wind produced by the solar wind interaction with Pluto’s atmosphere and both SWAP and PEPSSI can detect the energetic ions produced by the ionization of molecules from Pluto’s atmosphere.

### *4.3.2 Refine bulk parameters (radii, masses, densities) and orbits of Pluto and Charon*

The largest uncertainty in the heliocentric orbit of the Pluto system is its heliocentric distance. This will be improved by using Doppler ranging will determine the distance of the spacecraft, and optical navigation to determine the New Horizons trajectory relative to the Pluto system. Optical navigation is further aided by the differences in view angle to Pluto from New Horizons and the Earth.

LORRI will image the positions of Nix, Hydra, Charon, and Pluto over a 150-day timespan on approach, refining orbits of these bodies. Disk-resolved albedo maps will eliminate the persistent problem of center-of-light vs. center-of-mass offsets in Pluto-system astrometry (Buie et al. 2006). These orbits will improve measurements of the masses of Pluto and Charon, and possibly of Nix and Hydra. System masses will be augmented by REX Doppler measurements.

The radii and shapes of Pluto, Charon, Nix, and Hydra will be measured from imaging. Specifically, we will use LORRI for the highest geometric fidelity, obtaining best resolutions of order 3 km for all bodies (P_pan_global_1x1_last, C_pan_global_1x1_last). Alice and REX occultation measurements will augment measurements of the radii.

### *4.3.3 Search for magnetic fields of Pluto and Charon*

Being a small body it is unlikely that Pluto has a substantial region of molten, convection iron inside and hence unlikely to have a magnetic dynamo. Nevertheless, the weak interplanetary magnetic field and tenuous solar wind at >30 AU mean that should Pluto have a weak magnetic field then it would produce a relatively large magnetosphere. Bagenal et al. (1997) estimate that a magnetic field > 3700 nT would produce a magnetosphere that would hold off the solar wind upstream of the planet and produce a cavity (detected as a drop in solar wind flux by SWAP) extending beyond the flyby distance of the New Horizons spacecraft.

### *4.3.4 Search for additional satellites and rings.*

Since the time of the formulation of the measurement goals for New Horizons, two new satellites, Nix and Hydra, have been discovered (Weaver et al. 2006). The entire Hill sphere will be searched by LORRI and MVIC for additional rings and satellites at low solar phase angle during the approach phase. MVIC will search for rings in forward scattering about two hours after Pluto closest approach (P_pan_GlobalRings_scan_1). Finally, SDC will search for dust from the Pluto system in the departure phase.

# 5 Broader Impact

By deepening our understanding of the Pluto system in particular, New Horizons will extend our knowledge of planetary science in general. We mention just a few examples here.

The surface composition of Pluto presumably reflects the nature of the solar nebula at large heliocentric distances 4.6 billion years ago. Does the dominant role played by $N_2$ on Pluto's surface confirm the hypothesis that $N_2$ was the dominant nitrogen-bearing gas in the nebula? If $N_2$ was dominant in the nebula, why isn't that reflected in the composition of comets, which seem to be $N_2$ depleted? Are the $N_2$/CO and CO/$CH_4$ ratios on Pluto's surface consistent with current theories of the composition of the solar nebula? Solid $CO_2$ has been detected on Triton (Quirico et al 1999, Grundy and Young 2004), but not yet on Pluto or Charon. What will a detection or upper limit tell us about Pluto and Charon's original volatile budget, cosmic-radiation-drive surface chemistry, and geologic histories (e.g., Shock and McKinnon 1993, Grundy et al 2006)? Based on detailed composition measurements of Pluto and Charon, we will know more about the origin and evolution of comets and icy dwarfs in the outer solar system.

While the concept of a surface and atmosphere in vapor-pressure equilibrium is well established (Spencer et al 1997), many complications remain (Trafton et al. 1998). One is the relation between low-resolution albedo maps and volatile distribution. It is usually assumed that fresh ice is bright, and old ice, possibly chemically processed, is dark. However, new $N_2$ may be transparent if deposition rates are slow (Duxbury et al. 1997), or old frost might be brightened by the processes of subsidence of dark contaminants (sinking into pits created by their own local thermal heating; see Spencer et al. 1997). Moreover, the emissivity of $N_2$ frost, a factor in its equilibrium temperature, may depend on its crystalline phase (Stansberry and Yelle 1999). Another complication is the interaction of multi-component frosts with the atmosphere. On Pluto, the atmosphere is primarily $N_2$, with trace amounts of CO and $CH_4$, which are less volatile than $N_2$; all three species are seen as surface ices. Models for the multi-component interaction include the formation of a monolayer $CH_4$-rich cap above an $N_2$:$CH_4$ substrate, which chokes off communication of $N_2$ with the atmosphere, or the formation of warm, $CH_4$-rich patches, which leads to turbulent boundary layers that enhance $CH_4$ mixing ratios in the atmosphere. The powerful combinations of instruments on board New Horizons can investigate the relationship between geology, surface (and subsurface) energy balance, ice composition, mixing state, and temperature, along with the surface pressures and composition of the atmosphere. Of the eleven global atmospheres in our solar system, four are Jovian, three have atmospheres that do not condense on the surface, and four (Pluto, Triton, Io, and Mars) have surfaces volatiles with significant vapor pressures at the surface temperatures. Thus, study of the behavior of Pluto's surface-atmosphere interaction leads to a greater understanding of a broad classes of atmospheres. As we learn more about the surfaces of other icy dwarf planets, the knowledge from Pluto can extend our understanding of the possibilities of other seasonal atmospheres in the Kuiper belt.

The thermal structure and composition of Pluto's atmosphere, measured with Alice and REX, will serve as the basic input for models investigating atmospheric processes. These models will be further constrained by other New Horizons observations. The results of atmospheric chemistry may be detectable by the presence of condensation products on the surface, by hazes in the atmosphere, by their UV signatures, or by heavier ions detected by the particle instruments. Pluto, Triton, and Titan are often thought of as an atmospheric triad, since all three have $N_2$ dominated atmospheres with $CH_4$; an improved understanding of chemistry and energetics on Pluto will improve our understanding of all three. Furthermore, a deeper understanding of atmospheric escape on Pluto has application to the early Earth, whose atmosphere is also thought to have been in hydrodynamic escape.

The anticipated science results from the New Horizons mission presented here are based on our current level of knowledge of the Pluto system. New Horizons is the first mission to the Pluto system, and the first mission to an icy dwarf planet. We should be prepared to be surprised and amazed.

# Acknowledgments

The authors acknowledge the work of the cast of scientists who have studied the Pluto system, and the numerous NASA and NSF grants and telescope allocations that have supported this work. We are grateful to be working with each of the engineers on the New Horizons project. Finally, the authors thank the public and the scientific community alike for their sustained interest in Pluto and the Pluto system that has made this mission possible.